\documentstyle[epsfig,psfig]{mn}

\newif\ifAMStwofonts

\def\.{\'\i}

\def\etal{{\it et al. }}

\newcommand{\Nii}{[\hbox{N\,{\sc ii}}]\ }
\newcommand{\Oiii}{[\hbox{O\,{\sc iii}}]\ }
\ifoldfss
  \ifCUPmtlplainloaded \else
    \NewTextAlphabet{textbfit} {cmbxti10} {}
    \NewTextAlphabet{textbfss} {cmssbx10} {}
    \NewMathAlphabet{mathbfit} {cmbxti10} {} % for math mode
    \NewMathAlphabet{mathbfss} {cmssbx10} {} %  "   "    "
  \fi
  \ifAMStwofonts
    \ifCUPmtlplainloaded \else
      \NewSymbolFont{upmath} {eurm10}
      \NewSymbolFont{AMSa} {msam10}
      \NewMathSymbol{\upi}     {0}{upmath}{19}
      \NewMathSymbol{\umu}     {0}{upmath}{16}
      \NewMathSymbol{\upartial}{0}{upmath}{40}
      \NewMathSymbol{\leqslant}{3}{AMSa}{36}
      \NewMathSymbol{\geqslant}{3}{AMSa}{3E}

    \fi
  \fi
\fi % End of OFSS

\ifnfssone
  \newmathalphabet{\mathit}
  \addtoversion{normal}{\mathit}{cmr}{m}{it}
  \addtoversion{bold}{\mathit}{cmr}{bx}{it}
  \newmathalphabet{\mathbfit} % math mode version of \textbfit{..}
  \addtoversion{normal}{\mathbfit}{cmr}{bx}{it}
  \addtoversion{bold}{\mathbfit}{cmr}{bx}{it}
  \newmathalphabet{\mathbfss} % math mode version of \textbfss{..}
  \addtoversion{normal}{\mathbfss}{cmss}{bx}{n}
  \addtoversion{bold}{\mathbfss}{cmss}{bx}{n}
  \ifAMStwofonts
    \ifCUPmtlplainloaded \else
      \UseAMStwoboldmath
      \makeatletter
      \new@mathgroup\upmath@group
      \define@mathgroup\mv@normal\upmath@group{eur}{m}{n}
      \define@mathgroup\mv@bold\upmath@group{eur}{b}{n}
      \edef\UPM{\hexnumber\upmath@group}
      \new@mathgroup\amsa@group
      \define@mathgroup\mv@normal\amsa@group{msa}{m}{n}
      \define@mathgroup\mv@bold\amsa@group{msa}{m}{n}
      \edef\AMSa{\hexnumber\amsa@group}
      \makeatother
      \mathchardef\upi="0\UPM19
      \mathchardef\umu="0\UPM16
      \mathchardef\upartial="0\UPM40
      \mathchardef\leqslant="3\AMSa36
      \mathchardef\geqslant="3\AMSa3E
    \fi
  \fi
\fi % End of NFSS release 1

\ifnfsstwo
  \DeclareMathAlphabet{\mathbfit}{OT1}{cmr}{bx}{it}
  \SetMathAlphabet\mathbfit{bold}{OT1}{cmr}{bx}{it}
  \DeclareMathAlphabet{\mathbfss}{OT1}{cmss}{bx}{n}
  \SetMathAlphabet\mathbfss{bold}{OT1}{cmss}{bx}{n}
  \ifAMStwofonts
    \ifCUPmtlplainloaded \else
      \DeclareSymbolFont{UPM}{U}{eur}{m}{n}
      \SetSymbolFont{UPM}{bold}{U}{eur}{b}{n}
      \DeclareSymbolFont{AMSa}{U}{msa}{m}{n}
      \DeclareMathSymbol{\upi}{0}{UPM}{"19}
      \DeclareMathSymbol{\umu}{0}{UPM}{"16}
      \DeclareMathSymbol{\upartial}{0}{UPM}{"40}
      \DeclareMathSymbol{\leqslant}{3}{AMSa}{"36}
      \DeclareMathSymbol{\geqslant}{3}{AMSa}{"3E}
    \fi
  \fi
\fi % End of NFSS release 2

\ifCUPmtlplainloaded \else
  \ifAMStwofonts \else % If no AMS fonts
    \def\upi{\pi}
    \def\umu{\mu}
    \def\upartial{\partial}
  \fi
\fi

\title[II. Global trends]
 {Group, field and isolated early-type galaxies \\
  II. Global trends from nuclear data}

\author[G. Denicol\'o \etal]
    {G. Denicol\'o,$^1$\thanks{Visitor at INAOE, Mexico.} 
    Roberto Terlevich,$^2$\thanks{Visiting Fellow, IoA, Cambridge. E-mail for contact: {\it rjt@inaoep.mx}  \ and  \ {\it eterlevi@inaoep.mx}}  
    Elena Terlevich,$^2$\footnotemark[2]
    Duncan A. Forbes,$^3$ \cr
    Alejandro Terlevich$^4$ \\
$^1$Institute of Astronomy, Madingley Road, Cambridge, CB3 0HA, United Kingdom \\
$^2$Instituto Nacional de Astrof\'{\i}sica, \'Optica y Electr\'onica, 
Tonantzintla, Puebla, Mexico \\
$^3$Centre for Astrophysics \& Supercomputing, Swinburne University, Hawthorn, VIC 3122, Australia\\
$^4$School of Physics and Astronomy, University of Birmingham, Edgbaston, Birmingham, B15 2TT, United Kingdom}

\date{Accepted. Received; in original form 2004 July}

\pagerange{\pageref{firstpage}--\pageref{lastpage}}
\pubyear{2004}

\begin{document}

\maketitle

\label{firstpage}

%%%%%%%%%%%%%%%%%%%%%%%%%%%%%%%%%%%%%
%%%%%%%%%%%   ABSTRACT   %%%%%%%%%%%%
%%%%%%%%%%%%%%%%%%%%%%%%%%%%%%%%%%%%%

\begin{abstract}
We have derived ages, metallicities and enhanced-element ratios
[$\alpha$/Fe] for a sample of 83 early-type galaxies essentially
in groups, the field or isolated objects. The stellar population
properties derived for each galaxy corresponds to the nuclear
r$_e$/8 aperture extraction. The median age found for Es is 5.8
$\pm$ 0.6 Gyr and the average metallicity is +0.37 $\pm$ 0.03
dex. For S0s, the median age is 3.0 $\pm$ 0.6 Gyr and [Z/H] =
0.53 $\pm$ 0.04 dex.  We compare the distribution of our galaxies
in the H$\beta$-[MgFe] diagram with Fornax galaxies. Our
elliptical galaxies are 3-4 Gyr younger than Es in the Fornax
cluster. We find that the galaxies lie in a plane defined by
[Z/H] = 0.99 log $\sigma_0$ - 0.46 log {\it Age} - 1.60, or in
linear terms Z $\propto$ $\sigma_0$ $\cdot$ Age$^{-0.5}$. More
massive (larger $\sigma_0$) and older galaxies present, on
average, large [$\alpha$/Fe] values, and therefore, must have
undergone shorter star-formation timescales.  Comparing group
against field/isolated galaxies, it is not clear that environment
plays an important role in determining their stellar population history. In
particular, our isolated galaxies show ages differing in more
than 8 Gyr. Finally we explore our large spectral coverage to
derive log(O/H) metallicity from the H$\alpha$ and
\Nii$\lambda$6584 \AA\ and compare it with model-dependent [Z/H]. We
find that the O/H abundances are similar for all galaxies, and we
can interpret it as if most chemical evolution has already
finished in these galaxies.
\end{abstract}

%%%%%%%%%%%%%%%%%%%%%%%%%%%%%%%%%%%%%%%%%%%%%%%%%%%%%%%%%%%%%%%%%%%%%%%%%%%%%
%%%%%%%%%%%%%%%%%%%%%%%%%%%%%%%%%%%%%%%%%%%%%%%%%%%%%%%%%%%%%%%%%%%%%%%%%%%%%


%%%%%%%%%%%%%%%%%%%%%%%%%%%%%%%%%%%%%%%%%%%%%%
%%%%%%%%%%%%%%   INTRODUCTION   %%%%%%%%%%%%%%
%%%%%%%%%%%%%%%%%%%%%%%%%%%%%%%%%%%%%%%%%%%%%%
\section{Introduction}

A decision over the most appropriate scenario of formation and evolution of early-type galaxies is still an open question despite the enormous amount of work already put forward to model primary observational constraints and scaling relations. Most recent flavours of models ranging from monolithic collapse scenarios to hierarchical clustering predictions include Arimoto \& Yoshii (1987, 1989); Kauffmann, White \& Guiderdoni 1993; Bruzual \& Charlot (1993); Bressan, Chiosi \& Fagotto (1994); Worthey (1994); Einsel \etal (1995); Navarro \etal (1996); Steinmetz (1996); Tantalo \etal (1996); Bressan, Chiosi \& Tantalo (1996); Gibson (1996a,b, 1998); Gibson \& Matteucci (1997); Tantalo, Chiosi \& Bressan (1998); Chiosi \& Carraro (2002); Terlevich \& Forbes (2002); 
and references therein.

Briefly, in the conventional monolithic collapse scenario elliptical galaxies formed as a result of a single violent burst of star formation at high redshift (possibly z $>$ 3) and evolved quiescently ever since. By contrast, in the Hierarchical model, there is no intrinsic difference between star formation occurring at different epochs. All galaxies are viewed as similar star-forming systems in which there is an equilibrium between the inflow of gas and the rate at which it is either consumed or driven out of the galaxies by a supernova wind. Elliptical galaxies are formed by merger and/or accretion of smaller units over a time scale comparable to the Hubble time.

As pointed out by Trager \etal (2000b, hereafter T00b), the
evidence for intermediate-age stellar population
(between 1 and 10 Gyr) would favour hierarchical models, which
more naturally have extended star formation over time. On the
other hand, Willis et al. (2002) have presented an analysis of
the mean star formation history and space density of a sample of
luminous field (in their definition, field is equivalent to
non-cluster) early-type galaxies selected over the redshift
interval 0.3 $\la$ z $\la$ 0.6. In all, the mean star formation
history of the sample is characterised by an apparently old
(z$_{f}$ $>$ 1), solar to slightly above-solar metallicity
luminosity-weighted stellar population that has evolved passively
since the formation epoch. The mean properties of Willis et
al. sample are markedly similar to the properties of
morphologically-selected luminous elliptical galaxies in rich
cluster environments at redshifts z $<$ 1 (Ferreras \etal
1999). Though neither result in isolation constrains the extent
to which early-type galaxies in field or cluster environments
represent a co-eval or co-metal population, the broad similarity
between the star formation history of the dominant stellar mass
component in such galaxies is consistent with similar formation
conditions for each population.

Furthermore, the recent results from the K20 survey (Cimatti \etal 2002) have identified a class of old passively evolving objects (old-EROs) with derived minimum age of $\sim$ 3 Gyr, corresponding to a formation redshift of {\it z$_f$} $\ga$ 2.4. Pure luminosity evolution models with such formation redshifts well reproduce the density of old EROs (consistent with being passively evolving ellipticals), whereas the predictions of the current hierarchical merging models are lower than the observed densities by large factors (up to an order of magnitude). In this survey, a population of dusty star-forming galaxies is also identified at 0.7 $<$ {\it z} $<$ 1.5, equally populated as the old-EROs ensemble, and suggests a significant contribution to the cosmic star-formation density at {\it z$_f$} $\sim$ 1.

On the theoretical side, Chiosi \& Carraro (2002) nicely summarized the current situation by highlighting that the hierarchical scheme melting together subunits made of gas and stars is not the dominant one by which elliptical galaxies are made unless it has taken place in the very remote past.

Independent of the adopted scenario for the formation and
evolution of elliptical galaxies, models should meet some
observational constraints on the history of star-formation of
these objects. These constraints include ages, metallicities and
degree of enhancement in $\alpha$-elements, derived by means of
line-strength indices and Single Stellar Population (SSP) models
(e.g., Gonz\'alez 1993; Worthey 1994; Kuntschner 2000, 2001;
Trager \etal 2000a; T00b; Thomas, Maraston \& Bender 2003,
hereafter TMB03; Thomas, Maraston \& Korn 2004).

This paper explores the central stellar populations of a sample
of local early-type galaxies and search for correlations among
them and with structural parameters. Many previous works have
studied such correlations (e.g. Tantalo, Chiosi \& Bressan 1998;
J\o rgensen 1999; Trager \etal 2000a; T00b; Kuntschner 2000;
Terlevich \& Forbes 2002).  We
pay special attention to T00b, as many of the analysis presented
here were first explored by these authors.  We would like to
highlight the high mean signal-to-noise of our sample and the use
of the TMB03 set of SSP models for
the analysis in this paper.  TMB03 presented an empirical
calibration for the synthetic Lick indices of SSP models that for
the first time extended up to solar metallicity. The new
metallicity range covered approaches the regime that is relevant
for the interpretation of the integrated spectra of elliptical
galaxies.

In Section 2 we briefly describe the sample and stellar population model applied. In Section 3 we explore the line-strength Lick indices and the quality of the model calibrations. The determination of ages  and metallicities, and the enhanced-element ratio [$\alpha$/Fe] are presented in Sections 4 and 5, respectively. We investigate the existence of the age-metallicity-$\sigma_0$-[$\alpha$/Fe] hyperplane for early-type galaxies in Section 6. Discussions and conclusions from the nuclear stellar population analysis are presented in Sections 7 and 8. Finally, in Section 9 we explore our large spectral coverage to compare metallicities derived from different elements, in the form of a log(O/H) {\it vs} [Z/H] plot.

%%%%%%%%%%%%%%%%%%%%%%%%%%%%%%%%%%%%%%%%%%%%%%%%%%%%%%%%%%%%%%%%%%%%%%%%%%%%%
%%%%%%%%%%%%%%%%%%%%%%%%%%%%%%%%%%%%%%%%%%%%%%%%%%%%%%%%%%%%%%%%%%%%%%%%%%%%%

%%%%%%%%%%%%%%%%%%%%%%%%%%%%%%%%%%%%%%%%%%%%%%%%%%%%%%%%%%
%%%%%%% SECTION 2: DATA AND ADOPTED MODEL %%%%%%%%%%%%%%%
%%%%%%%%%%%%%%%%%%%%%%%%%%%%%%%%%%%%%%%%%%%%%%%%%%%%%%%%%%

\section{Data and adopted model}

In the present work we will be analysing the spectra for 83 galaxies with signal-to-noise (S/N) ratio greater than 15 (per resolution element FWHM = 6 \AA) for the r$_e$/8 central aperture extraction. The observations were performed at the 2.12 m telescope of the {\it Observatorio Astrof\'\i sico Guillermo Haro} (OAGH), at Cananea, Mexico. 
Eight galaxies in our sample were carefully classified as isolated (typical Tully 1987 density of 0.08); the sample also has 18 galaxies in the field, and 57 group members. The group classification is taken from Garcia (1993). The sample also contains three Virgo cluster galaxies, namely NGC~4365, NGC~4374, NGC~4754, with the respective Tully (1987) catalogue density of 2.93, 3.99 and 2.62. 
 In total, the sample consists of 52 elliptical galaxies and 31 bulges of S0s or early-type spirals. 
Note the original sample had 86 galaxies, but for the present analysis the galaxies NGC~3139, NGC~5854 and NGC~5869 were not used because they do not have spectral information at $\lambda$ $\la$ 5100 \AA, and so no ages and metallicities could be derived. Therefore, the total sample with age and [Z/H] information has 83 galaxies. The full details of the sample, observations, data reduction and index measurements are presented in Denicol\'o \etal (2004, hereafter Paper I).

The spectra cover a wavelength range from $\sim$ 3850 \AA\ to $\sim$ 6700 \AA. The median S/N of the sample is 40 per resolution element. The Lick/IDS line-strength index measurements were performed separately in each galaxy frame and later averaged, the error of the mean was computed from the repeated measurements per galaxy (on average, there are 8 spectral frames per galaxy). 

Kinematic information for the galaxies was carefully derived in Paper I. 25 Lick spectral indices were measured and calibrated to the Lick/IDS system. Most of the galaxies in the sample had the Balmer indices corrected for emission using H$\alpha$ detection. The emission correction with H$\alpha$ should be less uncertain than the empirical emission correction of the Balmer lines using \Oiii$\lambda$5007 \AA. 

In this work we refer to our sample as the OAGH sample, and use the following definitions for the special spectral indices: $<$Fe$>$ = (Fe4383+Fe5270+Fe5335)/3 and [MgFe] = $[Mgb\cdot(0.72 \cdot Fe5270 + 0.28 \cdot Fe5335)]^{1/2}$, as in Kuntschner (1998) and TMB03, respectively.

\

TMB03 have constructed stellar population models for various and
well-defined element abundance ratios. The most important ratio
is $\alpha$/Fe, which is the ratio of the so-called
$\alpha$-elements (O, Ne, Mg, Si, S, Ar, Ca) to the Fe-peak
elements (Cr, Mn, Fe, Co, Ni, Cu, Zn), because it carries
information on the formation time-scale of stellar populations.
The whole set of Lick indices of SSP models is presented with
[$\alpha$/Fe] ratios of 0.0, 0.3, 0.5.  The impact
from the element abundance changes on the absorption-line indices
are taken from Tripicco \& Bell (1995, TB95), using an extension
of the method introduced by T00b. The models are calibrated with
the globular cluster data of Puzia \etal (2002). The present
models now allow for the unambiguous derivation of SSP ages,
metallicities, and element abundance ratios, in particular
[$\alpha$/Fe] ratios. For this reason we have adopted their
models throughout this paper. We note that recent results from
the same group of authors (Thomas, Maraston \& Korn 2004) have
extended the models of variable abundance ratios to the Balmer
lines. The key result is that, unlike H$\beta$, the H$\delta$ and
H$\gamma$ indices show a marked dependence on [$\alpha$/Fe]
ratio. We will discuss this dependence again in the next Section.

%%%%%%%%%%%%%%%%%%%%%%%%%%%%%%%%%%%%%%%%%%%%%%%%%%%%%%%%%%
%%%%%%%%% SECTION 3: EXPLORING INDICES %%%%%%%%%%%%%%%%%%%
%%%%%%%%%%%%%%%%%%%%%%%%%%%%%%%%%%%%%%%%%%%%%%%%%%%%%%%%%%

\section{Exploring nuclear line-strength indices}

There are considerable shifts in metallicity and age when comparing 
TMB03 models with G. Worthey's SSP models (W94). 
 Figure 1 presents a comparison between  model tracks for 
the same age and [Fe/H] grid steps. 
Taking the position of the galaxy NGC~3379 in the diagram, we have interpolated an age of 10.9 $\pm^{2.6}_{2.8}$ Gyr and [Fe/H] = 0.30 $\pm^{0.07}_{0.08}$ dex using TMB03 models, while W94 models result in 11.4 $\pm^{3.9}_{2.6}$ Gyr and [Fe/H] = 0.22 $\pm^{0.07}_{0.08}$ dex. The ages and metallicities were interpolated as will be explained in Section 4.

The metallicity values derived from interpolation of the TMB03 model are 
generally higher than with the W94 model (in the case of NGC~3379, TMB03 
results +0.08 dex higher in metallicity than W94). Likewise, the ages are 
systematically younger in TMB03 by comparison with W94. 

We note that the emission correction (see Paper I, Section 4) used in this work is, in general, larger than in previous works. 
The effect of a higher emission correction, i.e., a higher H$\beta$ value, is a decrease in age and consequent increase in metallicity, due to the bending of the model tracks. Hence, one should keep in mind that both the
use of new model tracks and the larger emission 
correction are responsible for differences in age/metallicity values when comparing with the literature.
 
Note however, that the differences between the age and [Fe/H] from W94 and TMB03 models for NGC~3379 are not representative of the uncertainties involved in these determinations. The uncertainty in the age and [Fe/H] values for NGC~3379 is of greater magnitude than the difference between the models. We will discuss the errors in age and metallicity determinations in Section 4.1.

%%%%%%%%%%%%%%%%%%%%%%%%%%%%%%%%%%%%%%%%%%%%%%%%%%%%%%%%%%%%%%
%%%%%%%%%%%%%%%% FIGURE 1: MODELS COMPARISON %%%%%%%%%%%%%%%%%
%%%%%%%%%%%%%%%%%%%%%%%%%%%%%%%%%%%%%%%%%%%%%%%%%%%%%%%%%%%%%%
\begin{figure}
\vspace{8 cm}
\includegraphics{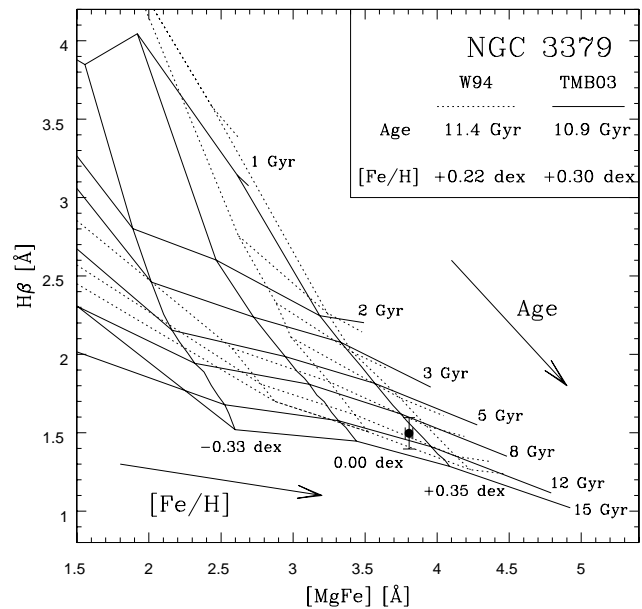}
\caption{Comparison between Worthey (1994, W94) and Thomas, Maraston \& Bender (2003, TMB03) models. Both models are shown in age steps of 1, 2, 3, 5, 8, 12, 15 Gyr and metallicity steps of -0.33, 0.00, +0.35. The Worthey model was interpolated to the given age and metallicity steps with the use of G. Worthey's home page (http://astro.wsu.edu/$\sim$worthey/). The point represents NGC 3379.}
\end{figure}
%%%%%%%%%%%%%%%%%%%%%%%%%%%%%%%%%%%%%%%%%%%%%%%%%%%%%%%%%%%%%%%

%%%%%%%%%%%%%%%%%%%%%%%%%%%%%
\subsection{Model vs data}

The aim of the initial paragraphs of this section is to decide which 
combination of indices provides a reliable and well-understood determination 
of the stellar population age and metallicity. For that, and to assess
the consistency of our index 
measurements  in relation to the model predictions we present 
index-index plots which are almost degenerate in age and metallicity. 
We separate the indices in three groups based upon their sensitivity to 
$\alpha$ elements, Fe-peak elements, and the Balmer lines. This group distinction is also useful to check how well the various Lick indices trace the corresponding element abundance. An analysis of the same type was carried out by TMB03, and we feel it is important to show the consistency of the models with respect to our sample as well.
In the following figures we plot 83 galaxies observed at OAGH.

%%%%%%%%%%%%%%%%%%%%%%%%%%%%%%%%%%%%%%%%%%%%%%%%%%%
%%%%%%%% FIGURE 2:  Index-index  [Balmer] %%%%%%%%%
%%%%%%%%%%%%%%%%%%%%%%%%%%%%%%%%%%%%%%%%%%%%%%%%%%%
\begin{figure}
\vspace{6.5 cm}
\includegraphics{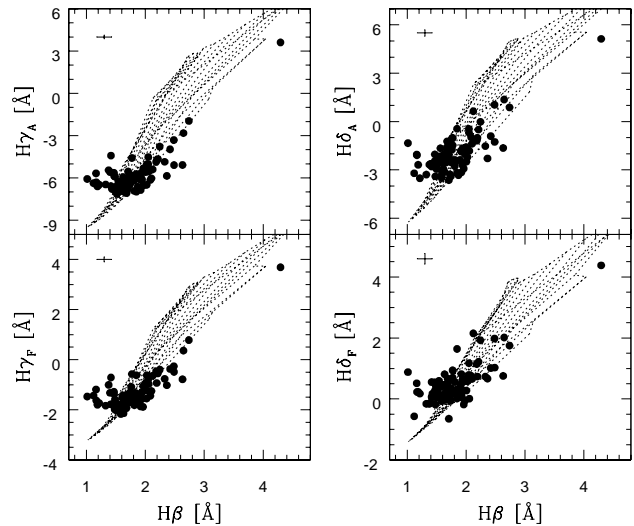}
\caption{Index {\it vs} index plots for Balmer-line indices in the Lick/IDS system. Overplotted are the model predictions by TMB03. The solid circles are 83 galaxies from our sample. The error bars represent the average error of the mean per index, from repeated observations.}
\end{figure}
%%%%%%%%%%%%%%%%%%%%%%%%%%%%%%%%%%%%%%%%%%%%%%%%%%%%

%%%%%%%%%%%%%%%%%%%%%%%%%%%%%%%%%%%%%%%%%%%%%%%%%%%%
%%%%%%%% FIGURE 3: Index-index [Alpha]   %%%%%%%%%%%
%%%%%%%%%%%%%%%%%%%%%%%%%%%%%%%%%%%%%%%%%%%%%%%%%%%%
\begin{figure}
\vspace{11 cm}
\includegraphics{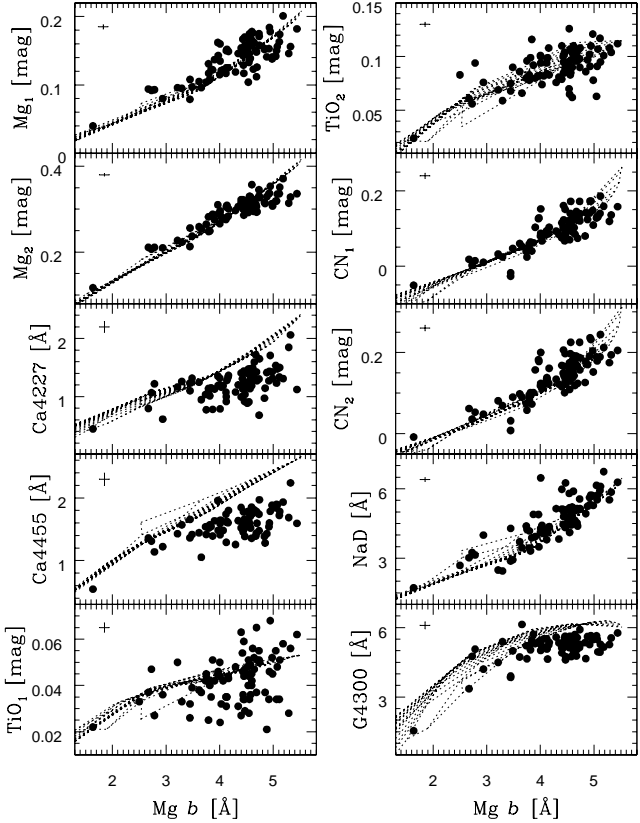}
\caption{Index {\it vs} index plots for indices that correlate with Mg. The symbols are the same as in Figure 2.}
\end{figure}
%%%%%%%%%%%%%%%%%%%%%%%%%%%%%%%%%%%%%%%%%%%%%%%%%%%%%

We analyse firstly the Balmer-line indices. Figure 2 shows the
relation between H$\beta$, H$\gamma_{A,F}$ and
H$\delta_{A,F}$. The H$\beta$ index is only marginally affected
by the [$\alpha$/Fe] ratio, as discussed in TB95 and in Thomas
\etal (2004), i.e., more important for the modelling is taking
correct account of the Horizontal Branch morphology (Maraston
\etal 2003). The models used here have the mass-loss on the red
giant branch phase taken into account, as calibrated in Maraston
\& Thomas (2000). The agreement between models and observations
is not as good as one would expect, even given the observational
errors. However, Thomas \etal (2004) discuss about the important
dependence of the higher order Balmer lines H$_{\delta,\gamma}$
on [$\alpha$/Fe]. To correctly compare the Balmer lines then we
would need to previously know the overabundance ratio for each
galaxy and use the corresponding model grid for each galaxy
[$\alpha$/Fe] value. This fact certainly diminishes the advantage
of higher-order Balmer line indices over H$\beta$ as age
indicators, as we will discuss in Section 4. We then emphasize to
the reader that some mismatch between galaxies and models of
higher-order Balmer lines at constant [$\alpha$/Fe] is expected,
because H$_{\delta,\gamma}$ change significantly for different
[$\alpha$/Fe] ratios.

In Figure 3 the $\alpha$-element sensitive indices Mg$_{1,2}$, TiO$_{1,2}$, NaD, CN$_{1,2}$ seem consistent with each other. 
The ``Ca lines'', Ca4227 and 
Ca4455, show a mismatch with the models; this mismatch grows with increasing Mg{\it b}. 
Indeed the model predictions of the Ca4227 index seem to be  
slightly too high as discussed in Maraston \etal (2003); 
those authors indicate that models with variable $\alpha$/Ca 
could explain the difference. 

We have plotted the Ca4455 index in the same diagram of the $\alpha$-element sensitive indices to exemplify how misleading can be the actual index nomenclature. Despite its name Ca4455 is insensitive to Ca abundance (TB95), while Fe and Cr, both elements of the ``depressed'' group, are the dominant contributors to this index. 
Concerning the model tracks, Ca4455  
responds strongly to [$\alpha$/Fe] ratio, and it most likely 
suffers from calibration problems (Maraston \etal 2003).

The TiO$_2$ agrees reasonably well with Mg{\it b} at different metallicities. 
 NaD also behaves consistently with Mg{\it b}, but the high sensitivity of the NaD index to interstellar absorption severely hampers its usefulness for stellar population studies (Worthey \etal 1994). We noticed that probably part of the scatter in the TiO$_1$ index is due to the broad NaD feature, once the blue pseudocontinuum of  TiO$_1$ is very close in wavelength to the NaD profile.
The CN indices show a good match between the galaxy sample and the models. 
We observe a mismatch of models and 
galaxy data for the G4300 index. The calibration of the models with the globular cluster data for this index is also not convincing as already noted by Puzia \etal (2002).

%%%%%%%%%%%%%%%%%%%%%%%%%%%%%%%%%%%%%%%%%%%%%%%
%%%%%%% FIGURE 4: Index-index  [Fe] %%%%%%%%%%%
%%%%%%%%%%%%%%%%%%%%%%%%%%%%%%%%%%%%%%%%%%%%%%%
\begin{figure}
\vspace{11 cm}
\includegraphics{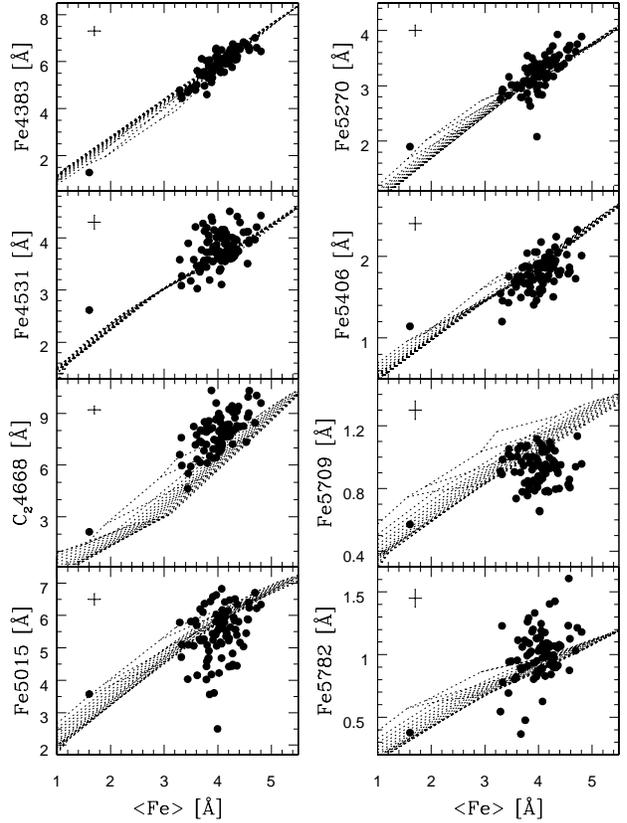}
\caption{Index {\it vs} index plots for Fe-peak indices. The symbols are the same as in Figure 2. }
\end{figure}
%%%%%%%%%%%%%%%%%%%%%%%%%%%%%%%%%%%%%%%%%%%%%%%

%%%%%%%%%%%%%%%%%%%%%%%%%%%%%%%%%%%%%%%%%%%%%%%%%%%%%%%%%%%%%%%
%%%%%%%       FIGURE 5:   BALMER x MgFe             %%%%%%%%%%%
%%%%%%%%%%%%%%%%%%%%%%%%%%%%%%%%%%%%%%%%%%%%%%%%%%%%%%%%%%%%%%%
\begin{figure*}
\vspace{21 cm}
\includegraphics{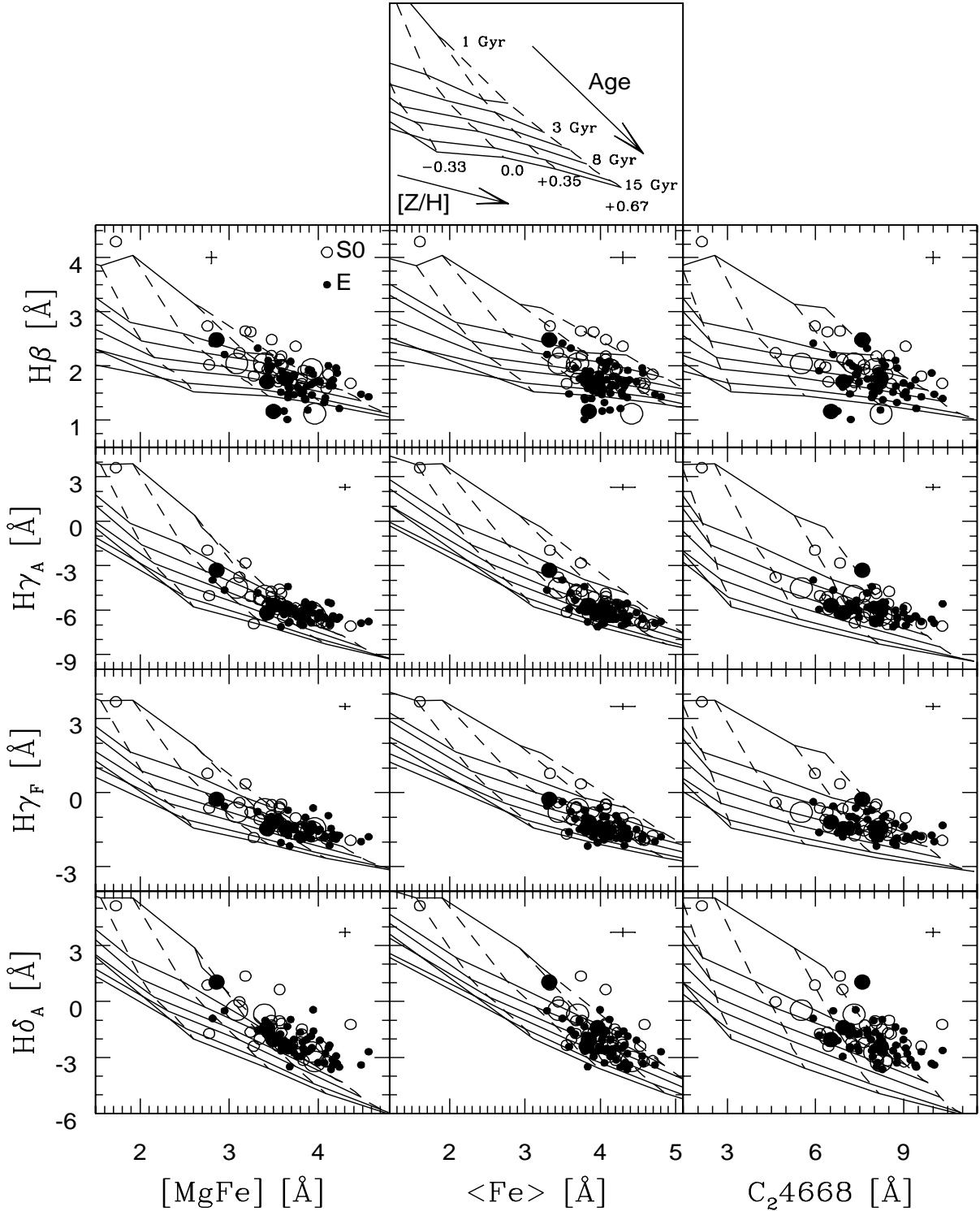}
\caption{Age-metallicity diagnostic diagrams. The Balmer-line indices are corrected for nebular emission as explained in Paper I. Overplotted are the model predictions by TMB03 (with [$\alpha$/Fe] = 0.0). The dashed lines represent lines of constant metallicity ([Z/H] = -1.35, -0.33, 0.00, +0.35, +0.67 dex), whereas the solid lines are lines of constant age
 (ages = 1, 2, 3, 5, 8, 12 and 15 Gyr). Open symbols denote S0 
galaxies, solid symbols represent elliptical galaxies. Isolated 
galaxies are plotted with larger symbols. The error 
bars at the top of the plots are the average error of the mean per index. The left-top corner S0 galaxy far away from the main distribution of galaxies is NGC~3156.}
\end{figure*}
%%%%%%%%%%%%%%%%%%%%%%%%%%%%%%%%%%%%%%%%%%%%%%%%%%%%%%%%%%%%%%%

In the diagrams with the indices most sensitive to Fe (Figure 4), we draw attention to the relatively large scatter in the index values for Fe5709, Fe5782, and specially for Fe5015, the last one 
mainly due to uncertainties in the emission contamination correction (the index Fe5015 was emission corrected as explained in Paper I, Section 
4.2.2). We plotted the C$_2$4668 index together with the Fe 
group as in Maraston \etal (2003). Although this index is supposed 
to follow carbon, it shows a correlation with Fe. C$_2$4668 
must have calibration problems too and its mismatch with the models cannot be explained by overabundance (TMB03). As in the globular clusters studies, this index shows a poor match between models and data (Puzia \etal 2002). In view of modelling uncertainties (TMB03; the mismatch of C$_2$4668 between model and observation cannot be explained only by $\alpha$/Fe effects), C$_2$4668 becomes an index not well suited for element abundance studies. 

In all, the distribution of the OAGH galaxies in these diagrams is very similar to the distribution of the bulk of galaxies from the sample of 381 Lick/IDS galaxies in Trager \etal (1998) (see comparison of data and models in TMB03). 

As pointed out by Worthey  (1994) the determination of the age of old 
stellar populations is complicated by the similar effects that age and 
metallicity have on the integrated spectral energy distributions. Broad 
band colours and most of the indices are degenerate along the locus of 
$\Delta$ age $\approx$ -3/2 $\Delta$ Z. Worthey dubbed this behaviour the 
``3/2 rule''. In other words, the integrated SED of an old ($\ga$ 2 Gyr) 
stellar population looks almost identical when age is doubled and Z reduced 
by a factor of three at the same time. Only a few narrow band weak 
absorption line indices have so far been identified which can potentially 
break this degeneracy. In terms of age, the Balmer lines H$\beta$, H$\gamma$ 
and H$\delta$ are the most promising features, being clearly more sensitive 
to age than to metallicity. Features which are more metal sensitive, and 
thus less age sensitive than the average are: C$_2$4668, Mg$_2$, Fe5270, 
Fe5335 and Fe4383.

In Figure 5 we show metallicity {\it vs} age indicators, in the form of [MgFe], $<$Fe$>$, C$_2$4668 against H$\beta$, H$\gamma_{A,F}$ and H$\delta_{A}$. Note that the degeneracy of the models is stronger for older metal rich systems,
i.e.~those in the bottom right corner of the diagrams.
Both ellipticals and S0s show 
a spread  in age of a factor of about 4 (mostly between 3-12 Gyr), and 
a larger spread in metallicities (approximately from -0.33 to +0.67 
dex). The outlier galaxy on the top-left corner of the plots in Figure 5 is 
NGC~3156,
a disk-dominated S0 galaxy with a rather faint bulge (Michard \& Marchal 
1994). 
As we have seen, when well calibrated by the models, C$_2$4668 
could be the most promising index for [Fe/H] determinations given its high 
sensitivity to metallicity (its range of variation from [Z/H] solar to three 
times solar is about 5 \AA). 
Considering ease of measurement, existing calibrations, large metallicity 
dependence, small age dependence, and relatively tight relation, we can say
that, at present, 
the best metallicity indicator in Figure 5 is [MgFe]. Note that we 
are using the [MgFe] index defined in TMB03, where Fe5270 and Fe5335 
are weighted as 
72\% and 28\%, respectively, to balance the [$\alpha$/Fe] dependence (see Fig.7 of TMB03). 
 
%%%%%%%%%%%%%%%%%%%%%%%%%%%%%%%%%%%%%%%%%%%%%%%%%%%%%%%%%%%%%
%%%%%%%       FIGURE 6:  Comparison with TF02     %%%%%%%%%%%
%%%%%%%%%%%%%%%%%%%%%%%%%%%%%%%%%%%%%%%%%%%%%%%%%%%%%%%%%%%%%
\begin{figure}
\vspace{12.5 cm}
%\special{psfile=./figures/testo.ps hoffset=-10 voffset=-60 hscale=70 vscale=70 angle=0}
\includegraphics{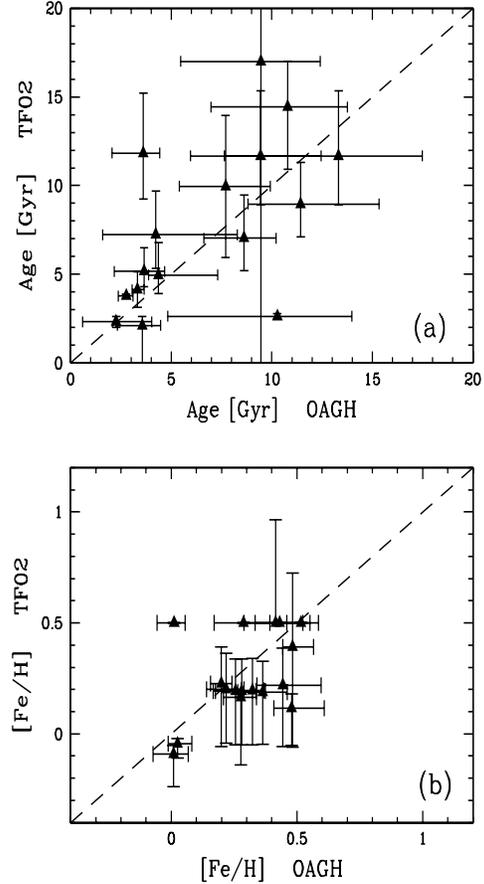}
\caption{The panels compare the ages and metallicities in the catalogue by Terlevich \& Forbes (2002, TF02) with our data determinations (OAGH).
One-to-one correspondence is indicated by the dashed line.
 We have interpolated ages and metallicities for 15 galaxies in common with TF02. Both age and metallicity values for this comparison were derived from the H$\beta$ {\it vs} [MgFe] diagram, using Worthey (1994) stellar population models (see Table 1).}
\end{figure}
%%%%%%%%%%%%%%%%%%%%%%%%%%%%%%%%%%%%%%%%%%%%%%%%%%%%%%%%%%%%%%%

Concerning age indicators, H$\beta$ is still the best calibrated 
Balmer-line index, even though it suffers from emission contamination. 
The higher order Balmer lines (i.e., H$\gamma$, H$\delta$) are a lot less sensitive to emission (see Table 6 in Paper I), but at the same time they are more affected by abundance  effects because of the presence of more metallic lines in the bluer parts of the spectrum (Thomas \etal 2004).
Maraston \etal (2003) present
a calibration of the Balmer lines, in H$_{\beta,\gamma,\delta}$ {\it vs} [MgFe] diagrams, 
for 12 globular clusters with [MgFe] $\la$ 3.5 \AA.  We emphasize
that, as in TMB03 (their Fig.2, concerning index calibration) the
behaviour of globular clusters does not always represent the range of values covered by galaxies. 
There are significant differences in the index calibration for clusters and galaxies specially for indices such as C$_2$4668, CN$_{1,2}$, Ca4227, Ca4455, NaD, TiO$_{1,2}$, H$\delta$ and H$\gamma$. The H$\beta$-index calibration for globular clusters seems
 robust, despite the very large scatter of the galaxy sample from Trager \etal (1998). 

%%%%%%%%%%%%%%%%%%%%%%%%%%%%%%%%%%%%%%%%%%%%%%%%%%%%%%%%%%%%%%%
%%%%%%%%%%%%%%%% TABLE 1 : Comparison with TF02 %%%%%%%%%%%%%%%
%%%%%%%%%%%%%%%%%%%%%%%%%%%%%%%%%%%%%%%%%%%%%%%%%%%%%%%%%%%%%%%
\begin{table}
\scriptsize
\centering
\begin{minipage}{85mm}
\caption{Comparison with TF02 of ages and metallicities derived using Worthey (1994) models.}
\vskip 0.2cm
\begin{tabular}{ l@{\hspace{.6em}} r@{\hspace{.5em}} r@{\hspace{.4em}} r@{\hspace{.0em}} r@{\hspace{.0em}}  }
\hline
\multicolumn{1}{c}{Galaxy} &
\multicolumn{1}{c}{Age OAGH} &
\multicolumn{1}{c}{Age TF02} &
\multicolumn{1}{c}{[Fe/H] OAGH}&
\multicolumn{1}{c}{[Fe/H] TF02} \\
\multicolumn{1}{c}{} &
\multicolumn{1}{c}{[Gyr]} &
\multicolumn{1}{c}{[Gyr]} &
\multicolumn{1}{c}{[dex]} &
\multicolumn{1}{c}{[dex]} \\
\hline      
NGC~221     &  2.8 $^{+0.3}_{-0.4}$ &  3.8 $^{+0.1}_{-0.1}$ &  0.03 $^{+0.06}_{-0.04}$ & -0.04 $^{+0.06}_{-0.02}$ \\
NGC~584     &  3.6 $^{+0.9}_{-1.2}$ &  2.1 $^{+0.5}_{-2.1}$ &  0.42 $^{+0.11}_{-0.08}$ &  0.50 $^{+0.46}_{-0.0}$  \\
NGC~821     &  4.2 $^{+4.0}_{-2.6}$ &  7.2 $^{+2.5}_{-1.9}$ &  0.44 $^{+0.15}_{-0.10}$ &  0.22 $^{+0.17}_{-0.28}$ \\
NGC~1700    &  2.3 $^{+1.8}_{-1.6}$ &  2.3 $^{+0.3}_{-0.3}$ &  0.43 $^{+0.15}_{-0.14}$ &  0.50 $^{+0.00}_{-0.00}$   \\
NGC~2300    &  4.4 $^{+2.9}_{-0.5}$ &  5.0 $^{+1.8}_{-1.1}$ &  0.48 $^{+0.08}_{-0.04}$ &  0.39 $^{+0.33}_{-0.45}$ \\
NGC~3377    &  3.3 $^{+0.3}_{-0.3}$ &  4.1 $^{+1.0}_{-1.0}$ &  0.32 $^{+0.05}_{-0.04}$ &  0.20 $^{+0.14}_{-0.24}$ \\
NGC~3379    & 11.4 $^{+3.9}_{-2.6}$ &  9.0 $^{+2.3}_{-1.9}$ &  0.22 $^{+0.07}_{-0.08}$ &  0.20 $^{+0.16}_{-0.24}$ \\
NGC~3608    &  7.7 $^{+2.2}_{-2.3}$ & 10.0 $^{+4.0}_{-4.0}$ &  0.28 $^{+0.06}_{-0.07}$ &  0.16 $^{+0.02}_{-0.30}$ \\
NGC~4261    & 10.8 $^{+3.0}_{-3.8}$ & 14.4 $^{+2.6}_{-3.5}$ &  0.36 $^{+0.10}_{-0.10}$ &  0.19 $^{+0.14}_{-0.23}$ \\ 
NGC~4374    &  3.6 $^{+0.8}_{-1.6}$ & 11.8 $^{+3.4}_{-2.6}$ &  0.48 $^{+0.13}_{-0.07}$ &  0.12 $^{+0.06}_{-0.17}$ \\
NGC~5638    &  8.6 $^{+1.6}_{-2.0}$ &  7.0 $^{+2.4}_{-1.8}$ &  0.20 $^{+0.04}_{-0.04}$ &  0.23 $^{+0.17}_{-0.28}$ \\
NGC~5831    & 10.3 $^{+3.7}_{-5.4}$ &  2.6 $^{+0.2}_{-0.2}$ &  0.29 $^{+0.10}_{-0.12}$ &  0.50 $^{+0.00}_{-0.00}$ \\
NGC~5846    &  9.4 $^{+3.0}_{-3.5}$ & 11.7 $^{+3.7}_{-2.7}$ &  0.26 $^{+0.08}_{-0.09}$ &  0.19 $^{+0.14}_{-0.24}$ \\
NGC~7454    &  3.6 $^{+1.0}_{-1.5}$ &  5.2 $^{+1.3}_{-0.9}$ &  0.01 $^{+0.06}_{-0.08}$ & -0.09 $^{+0.15}_{-0.04}$ \\
NGC~7626    & 13.3 $^{+4.2}_{-5.7}$ & 11.7 $^{+3.6}_{-2.7}$ &  0.28 $^{+0.09}_{-0.10}$ &  0.19 $^{+0.14}_{-0.24}$ \\

\hline
\end{tabular}
\end{minipage}
\end{table}

%%%%%%%%%%%%%%%%%%%%%%%%%%%%%%%%%%%%%%%%%%%%%%%%%%%%%%%%%%%%%%%%%%%%%%%%%

\

In this work, ages and metallicities (according to the TMB03 models)
are derived from the [MgFe] {\it vs} H$\beta$ plane as shown in Figure
5 (and  also in the top panel of Figure 7). Note that there is some
level of mismatch between data and models and a significant number of
galaxies falls outside the models.  We interpolated between the model
grid points, but did not attempt to extrapolate to ages or
metallicities greater than 15 Gyr or 0.67 dex, respectively (in the
plots we have denoted these galaxies by arrows and placed them with
symbolic values for age: 16 Gyr and [Z/H]: 0.69 dex). The only galaxy
we attempted to extrapolate a consistent age and [Z/H] value was
NGC~3156.  We have used a polynomial interpolation of varying order,
depending on the position of the galaxy in the [MgFe] {\it vs}
H$\beta$ diagram. For instance, galaxies placed in the region of 1-3
Gyr have less neighbouring model points  (i.e., lower order) than
galaxies older than 8 Gyr (more neighbouring data  points = higher
order interpolation). This is because, as we have stated before, the
degeneracy of the models is stronger for older systems.  Every
determination of age and metallicity was double-checked against the
location in the [MgFe] {\it vs} H$\beta$ plane for consistency of the
results.  We should always bear in mind that non-linear
interpolations using  distant spaced model points as in the case of
TMB03 model grids, may result  in non-stable determinations and there
is always the possibility of wild  oscillations between the tabulated
points (see discussion about interpolation  algorithms in Press \etal
1992).

It is important to notice that the {\it ages and metallicities derived here should not be taken as absolute values}; there are systematic offsets if different index combinations are used, although the relative ranking of the galaxies in the sample should not be very different (cf. Figure 5). In the TMB03 system, the average age for the group, field and isolated sample (excluding the three Virgo cluster galaxies) is 4.2 Gyr, with an average metallicity of +0.41 dex. The cluster galaxies in our sample also present rather young ages (NGC~4365: 3.6 Gyr, NGC~4374: 3.8 Gyr, NGC~4754: 5.8 Gyr). Both elliptical and S0 galaxies in our sample show similar spread in age and metallicity in the [MgFe]-H$\beta$ diagram with a marginal tendency for the bulge of S0s to be slightly younger than the elliptical galaxies, as will be shown in Section 4.

\

In panels {\it a} and {\it b} of Figure 6 we compare our determinations with ages and metallicities in the Terlevich \& Forbes (2002, TF02) catalogue, for which the authors have compiled ages, metallicities and abundances from the highest quality data of early-type galaxies in the literature. In TF02, ages and [Fe/H] were derived by interpolating Worthey (1994, W94) stellar population models in the H$\beta$-[MgFe] plane. To carry out this comparison, we have also interpolated ages and metallicities using W94 models.

Figure 6 shows a reasonably good agreement, inside the errors, between 15 galaxies in common with TF02 (see also Table 1). Our ages are in general slightly younger most probably due to the emission correction applied in this work. 
Panel {\it b} reveals the corresponding small shift to higher metallicities also very likely caused by the higher H$\beta$ values resulting after the emission correction.
There is indeed a correlation between $\Delta$ age (TF02-OAGH) and the
applied emission correction EC(H$\beta$) (Paper I). 

%%%%%%%%%%%%%%%%%%%%%%%%%%%%%%%%%%%%%%%%%%%%%%%%%%%%%%%%%%%%%%%

%%%%%%%%%%%%%%%%%%%%%%%%%%%%%%%%%%%%%%%%%%%%%%%%%%%%%%%%%%%%%%%%%%%
%%%%%%%%%% SECTION 4: AGES AND METALLICITIES %%%%%%%%%%%%%%%%%%%%%%
%%%%%%%%%%%%%%%%%%%%%%%%%%%%%%%%%%%%%%%%%%%%%%%%%%%%%%%%%%%%%%%%%%%

\section{Ages and metallicities}

%%%%%%%%%%%%%%%%%%%%%%%%%%%%%%%%%%%%%%%%%%%%%%%%%%%%%%%%%%%%%%%
%%%%%%%%%%%%% FIGURE 7: AGE X METALLICITY %%%%%%%%%%%%%%%%%%%%%
%%%%%%%%%%%%%%%%%%%%%%%%%%%%%%%%%%%%%%%%%%%%%%%%%%%%%%%%%%%%%%%
\begin{figure*}
\vspace{19 cm}
\includegraphics{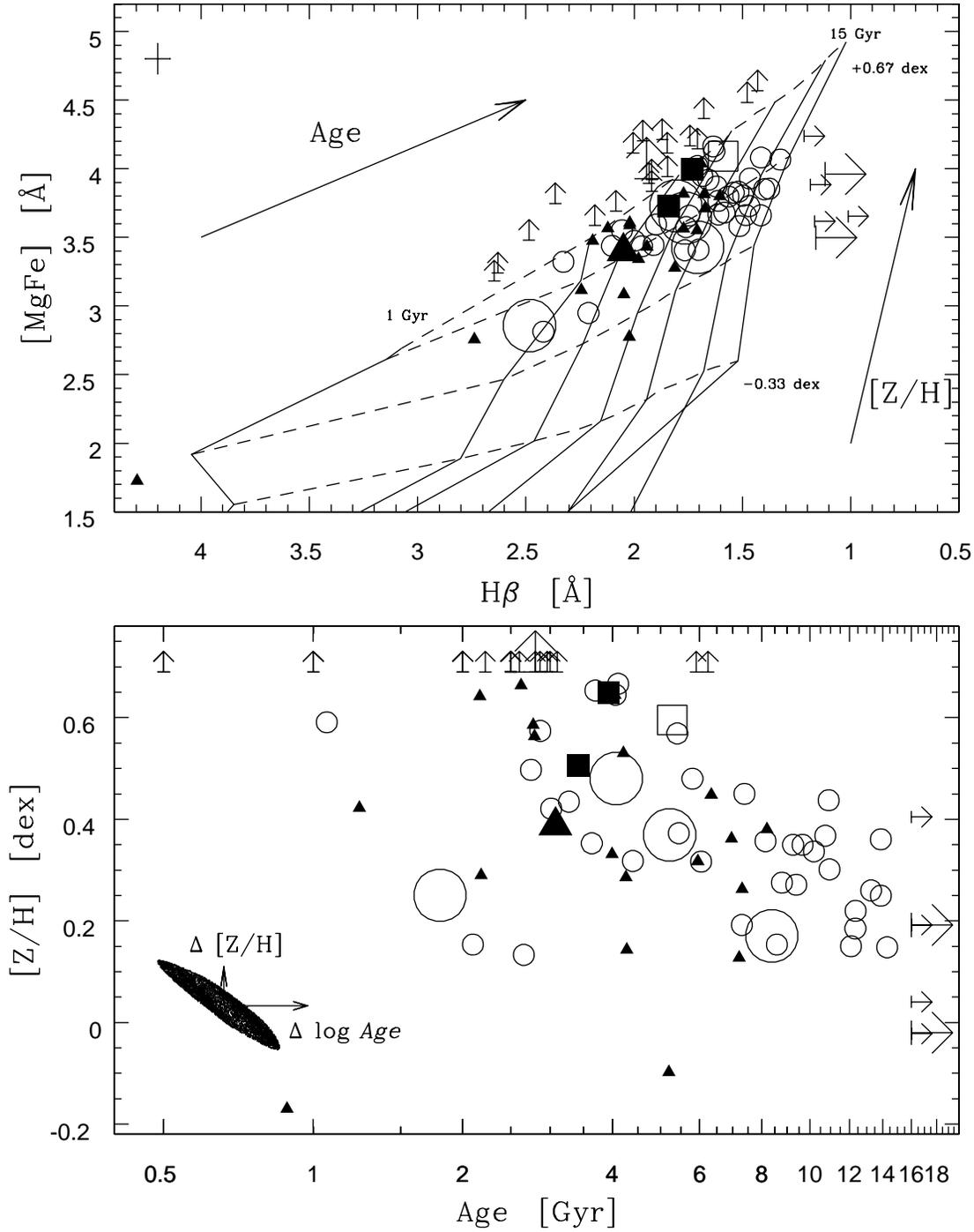}
\caption{The top panel shows the H$\beta$-[MgFe] plane from where Age and [Z/H] were interpolated using the TMB03 model grids. The model grids are shown in age steps of 1, 2, 3, 5, 8, 12, 15 Gyr and metallicity steps of -0.33, 0.00, +0.35, +0.67 dex.
The derived age {\it vs} [Z/H] diagram is shown in the bottom panel. Open circles represent elliptical galaxies and filled triangles are S0-type galaxies in the sample. Large circles and the large triangle are isolated galaxies. The three Virgo cluster galaxies in the sample, NGC~4365, NGC~4374, NGC~4754, are plotted as square symbols (S0 is open square, Es are filled squares). The average 1$\sigma$ error ellipse is presented at the bottom left corner of the age {\it vs} [Z/H] diagram. The vectors shown in the error ellipse correspond to 0.078 dex in [Z/H] and 0.17 in log Age.}
\end{figure*}
%%%%%%%%%%%%%%%%%%%%%%%%%%%%%%%%%%%%%%%%%%%%%%%%%%%%%%%%%%%%%%%

The [Z/H] -- log {\it Age} relation is presented in the bottom panel of 
Figure 7. The small arrows represent lower limits in age and metallicity for 
the galaxies that fall outside the model grids. Those galaxies for which we 
can only
give a lower limit in age are mostly ellipticals, while those for which we
give a lower limit in metallicity are as many ellipticals as S0s.
It is worth emphasizing that the age and metallicity values derived are not absolute, only the ranking of the galaxies is relatively independent of the adopted models. 
An error ellipse is shown at the bottom corner of the plot. This 1$\sigma$ error ellipse is the result of the interpolation in the H$\beta$-[MgFe] plane of a galaxy with H$\beta$ error of 0.11 \AA\ and [MgFe] error of 0.06 \AA, which correspond to the mean error values for these indices in the sample. Note that the correlation between age and [Z/H] in Figure 7 (roughly $\Delta$ log {\it age} $\cong$ $-$1.57 $\Delta$ [Z/H]) is consistent with being originated by correlated errors, as the spread of points has similar orientation to the major-axis of the error ``ellipse". A Spearman correlation test results in $\rho$ = -0.566 with more than 90\% probability that a correlation is present. 
The vectors shown in the error ellipse correspond to 0.078 dex in [Z/H] and 0.17 dex in log Age; these values are a rough average of the overall uncertainties in age and metallicity determinations for our sample.
A further analysis of the errors in the age-metallicity diagram is presented in the next Section.

S0s and Es show a slightly different distribution in the plane
[Z/H] {\it vs} log {\it Age}. Excluding the three Virgo cluster
galaxies, the sample of mostly low-density environment galaxies
is composed of 50 ellipticals and 30 S0s. The median age value
for Es is 5.8 $\pm$ 0.6 Gyr, and for S0s is 3.0 $\pm$ 0.6 Gyr,
where the uncertainties are the errors of the mean. The median
metallicity value for Es is 0.37 $\pm$ 0.03 dex and for S0s, 0.53
$\pm$ 0.04 dex. Therefore S0 galaxies seem to be slightly younger
and more metal-rich than ellipticals. However the elliptical
galaxies in our sample show a larger spread in age distribution
than S0s.  Note also that the metallicities of the young S0
galaxies are subject to large errors because they are likely to
have disk contamination, i.e., composite stellar populations.  In
particular, the isolated galaxies themselves also cover a large
range in metallicity {\it and} age.  On the other hand, our two
Virgo cluster elliptical galaxies are placed relatively close to
each other in age (filled square symbols in Figure 7). It might
be expected that cluster galaxies constitute a more homogeneous
sample (as demonstrated in the Fornax cluster by Kuntschner
1998), with perhaps a larger spread in metallicities than in
ages. Table 2 presents the ages and [Z/H] values for our sample.

It is important to note that the age/metallicity analysis presented here
depends on the assumption of a single (luminosity weighted) stellar 
population, which we feel is a reasonable one. The presence of additional
populations (especially if they are luminous) will bias -- perhaps 
significantly -- the resulting age/metallicity estimates.
(Bressan \etal 1996; Trager \etal 2000a; T00b).

%%%%%%%%%%%%%%%%%%%%%%%%%%%%%%%%%%%%%%%%%%%%%%%%%%%%%%%%%%%%%%%%
%%%%%%%%%%%%%% TABLE 2:  AGES AND METALLICITIES %%%%%%%%%%%%%%%%
%%%%%%%%%%%%%%%%%%%%%%%%%%%%%%%%%%%%%%%%%%%%%%%%%%%%%%%%%%%%%%%%
\begin{table*}
\caption{Age, [Z/H] and [$\alpha$/Fe] from nuclear r$_e$/8 extractions.}
\vspace{0.2 cm}
\centering
\begin{minipage}{180mm}
\begin{tabular}{@{}lrrrclrrr@{}}
\hline
\multicolumn{1}{c}{Galaxy} &
\multicolumn{1}{c}{Age} &
\multicolumn{1}{c}{[Z/H]} &
\multicolumn{1}{c}{[$\alpha$/Fe]$^{\dagger}$} &
\multicolumn{1}{c}{$\vline$} &
\multicolumn{1}{c}{Galaxy} &
\multicolumn{1}{c}{Age} &
\multicolumn{1}{c}{[Z/H]} &
\multicolumn{1}{c}{[$\alpha$/Fe]$^{\dagger}$} \\
\multicolumn{1}{c}{} &
\multicolumn{1}{c}{[Gyr]} &
\multicolumn{1}{c}{[dex]} &
\multicolumn{1}{c}{[dex]} &
\multicolumn{1}{c}{$\vline$} &
\multicolumn{1}{c}{} &
\multicolumn{1}{c}{[Gyr]} &
\multicolumn{1}{c}{[dex]} &
\multicolumn{1}{r}{[dex]} \\
\hline      
{\bf  ESO462-G015}   &  5.6 $^{+1.0}_{-2.0}$ &  0.37 $^{+0.17}_{-0.20}$ & 0.24 & $\vline$ & NGC 3414      &  6.0 $^{+3.0}_{-3.7}$ &  0.43 $^{+0.09}_{-0.13}$ & 0.34  \\
MCG-01-03-018 & $>$ $^{+1.2}_{-3.6}$ &  0.19 $^{+0.06}_{-0.06}$ & 0.09 & $\vline$ & NGC 3599  &  1.2 $^{+0.1}_{-0.4}$ & 0.30 $^{+0.15}_{-0.14}$ &-0.04 \\
NGC 0016      &  2.8 $^{+0.05}_{-0.1}$ &  0.56 $^{+0.04}_{-0.05}$ & 0.10  & $\vline$ & NGC 3607  &  2.7 $^{+1.0}_{-0.7}$ & 0.67$>$ $^{+0.12}_{-0.07}$ & 0.17 \\
{\bf  NGC 0221}       &  2.1 $^{+0.2}_{-0.2}$ &  0.15 $^{+0.04}_{-0.05}$ &-0.10  & $\vline$ & {\bf  NGC 3608}  &  8.1 $^{+2.8}_{-3.1}$ & 0.36 $^{+0.04}_{-0.17}$ & 0.27 \\
{\bf  NGC 0315}       &  2.5 $^{+1.0}_{-1.0}$ &  $>$ $^{+0.37}_{-0.02}$ & 0.20  & $\vline$ & {\bf   NGC 3610}  &  1.8 $^{+0.2}_{-0.5}$ & 0.59 $^{+0.21}_{-0.18}$ & 0.12 \\
NGC 0474       & 7.3 $^{+2.0}_{-2.4}$ &  0.26 $^{+0.07}_{-0.07}$ & 0.09 & $\vline$ &  {\bf  NGC 3613}  &  5.6 $^{+1.7}_{-2.5}$ & 0.37 $^{+0.07}_{-0.07}$ & 0.10 \\
{\bf  NGC 0584}       &  3.8 $^{+0.5}_{-1.7}$ &  0.43 $^{+0.07}_{-0.10}$ & 0.15  & $\vline$ & {\bf  NGC 3636}  &  4.4 $^{+1.4}_{-2.0}$ & 0.32 $^{+0.09}_{-0.10}$ & 0.08 \\ 
{\bf  NGC 0720}       &  3.0 $^{+0.5}_{-4.3}$ &  $>$ $^{+0.12}_{-0.02}$ & 0.23  & $\vline$ & {\bf  NGC 3640}  &  2.5 $^{+0.7}_{-0.3}$ & 0.53 $^{+0.13}_{-0.13}$ & 0.12 \\
{\bf  NGC 0750}       & 14.0 $^{+3.4}_{-1.7}$ &  0.25 $^{+0.04}_{-0.10}$ & 0.23  & $\vline$ & NGC 3665  &  2.0 $^{+1.0}_{-0.8}$ & $>$ $^{+0.08}_{-0.07}$ & 0.16 \\
{\bf  NGC 0751}       &  8.6 $^{+1.9}_{-2.4}$ &  0.15 $^{+0.08}_{-0.08}$ & 0.03 & $\vline$ & {\bf  NGC 3923}  &  2.6 $^{+0.5}_{-0.6}$ & $>$ $^{+0.04}_{-0.07}$ & 0.14 \\
{\bf  NGC 0777}       & 13.3 $^{+6.7}_{-5.2}$ &  0.26 $^{+0.20}_{-0.20}$ & 0.26  & $\vline$ & NGC 3941  &  2.6 $^{+0.4}_{-0.3}$ & 0.55 $^{+0.13}_{-0.12}$ & 0.07 \\
{\bf  NGC 0821}       &  4.0 $^{+1.7}_{-3.5}$ &  0.48 $^{+0.13}_{-0.17}$ & 0.17 & $\vline$ & {\bf  NGC 4125}  &  5.9 $^{+2.5}_{-3.0}$ & 0.32 $^{+0.10}_{-0.10}$ & 0.10 \\
NGC 0890       &  2.2 $^{+0.4}_{-0.4}$ &  0.60 $^{+0.08}_{-0.07}$ & 0.12  & $\vline$ & {\bf  NGC 4261}  & 10.6 $^{+4.7}_{-4.7}$ & 0.44 $^{+0.10}_{-0.03}$ & 0.22 \\
NGC 1045      &  2.6 $^{+0.8}_{-2.7}$ &  $>$ $^{+0.14}_{-0.06}$ & 0.14  & $\vline$ & {\bf  NGC 4365}  &  3.6 $^{+3.7}_{-2.3}$ & 0.65 $^{+0.10}_{-0.06}$ & 0.20 \\
{\bf  NGC 1052}      &  2.9 $^{+0.4}_{-8.8}$ &  $>$ $^{+0.34}_{-0.03}$ & 0.34  & $\vline$ & {\bf  NGC 4374}  &  3.8 $^{+1.1}_{-2.7}$ & 0.50 $^{+0.11}_{-0.15}$ & 0.23 \\
{\bf  NGC 1132}      & $>$ $^{+5.1}_{-14.9}$ &  -0.02 $^{+0.19}_{-0.16}$ & 0.24  & $\vline$ & {\bf  NGC 4494}  &  6.7 $^{+0.4}_{-0.3}$ & 0.19 $^{+0.04}_{-0.04}$ & 0.17 \\
{\bf  NGC 1407}      &  2.5 $^{+0.5}_{-0.8}$ &  $>$ $^{+0.01}_{-0.09}$ & 0.18  & $\vline$ & NGC 4550  &  2.3 $^{+1.4}_{-4.4}$ & 0.29 $^{+0.25}_{-0.40}$ & 0.05 \\
{\bf  NGC 1453}      & $>$ $^{+4.7}_{-1.7}$ &  0.45 $^{+0.04}_{-0.09}$ & 0.19  & $\vline$ & NGC 4754  &  5.8 $^{+1.9}_{-2.5}$ & 0.55 $^{+0.07}_{-0.08}$ & 0.15 \\
{\bf  NGC 1600}      &  2.7 $^{+0.5}_{-1.0}$ &  $>$ $^{+0.00}_{-0.06}$ & 0.13 & $\vline$ & {\bf  NGC 5322}  &  2.4 $^{+0.4}_{-0.4}$ & $>$ $^{+0.09}_{-0.06}$ & 0.14 \\
{\bf  NGC 1700}      &  2.6 $^{+1.4}_{-1.8}$ &  0.45 $^{+0.19}_{-0.21}$ & 0.11  & $\vline$ & NGC 5353  &  3.0 $^{+0.3}_{-1.7}$ & $>$ $^{+0.02}_{-0.03}$ & 0.21 \\
NGC 1726      &  7.0 $^{+4.0}_{-5.4}$ &  0.36 $^{+0.07}_{-0.16}$ & 0.34  & $\vline$ & {\bf  NGC 5354} 90$^o$  & $>$ $^{+0.9}_{-7.3}$ & 0.00 $^{+0.09}_{-0.05}$ & 0.14 \\
NGC 2128      &  3.1 $^{+2.6}_{-5.4}$ &  0.39 $^{+0.16}_{-0.36}$ & 0.30 & $\vline$ & {\bf  NGC 5354} 0$^o$ & 9.3 $^{+3.6}_{-3.9}$ & 0.34 $^{+0.13}_{-0.06}$ & 0.14 \\
NGC 2300      &  4.0 $^{+1.2}_{-1.4}$ &  0.53 $^{+0.05}_{-0.06}$ & 0.19  & $\vline$ & NGC 5363  &  3.8 $^{+2.1}_{-3.5}$ & 0.29 $^{+0.11}_{-0.19}$ & 0.24 \\
{\bf  NGC 2418}      &  4.8 $^{+2.6}_{-2.8}$ &  0.57 $^{+0.12}_{-0.11}$ & 0.16 & $\vline$ & {\bf  NGC 5444}  & $>$ $^{+0.5}_{-3.4}$ & 0.17 $^{+0.05}_{-0.03}$ & 0.25 \\
{\bf  NGC 2513}      &  3.8 $^{+2.4}_{-2.0}$ &   $^{+0.07}_{-0.08}$ & 0.22 & $\vline$ & {\bf  NGC 5557}  &  7.0 $^{+1.3}_{-1.4}$ & 0.45 $^{+0.07}_{-0.07}$ & 0.18 \\
NGC 2549      &  1.5 $^{+0.3}_{-0.5}$ &  $>$ $^{+0.01}_{-0.20}$ & 0.05 & $\vline$ & {\bf  NGC 5576}  &  3.2 $^{+0.2}_{-1.2}$ & 0.42 $^{+0.08}_{-0.11}$ & 0.07 \\
NGC 2768 0$^o$  &  8.0 $^{+3.5}_{-3.8}$ &  0.38 $^{+0.29}_{-0.20}$ & 0.12  & $\vline$ & {\bf  NGC 5638}  &  8.8 $^{+1.4}_{-1.5}$ & 0.28 $^{+0.05}_{-0.05}$ & 0.21 \\
NGC 2768 90$^o$ &  3.8 $^{+1.0}_{-1.3}$ &  0.64 $^{+0.06}_{-0.03}$ & 0.12  & $\vline$ & {\bf  NGC 5812}  &  3.8 $^{+0.8}_{-1.6}$ & 0.65 $^{+0.15}_{-0.03}$ & 0.21 \\
{\bf  NGC 2872}      &  9.7 $^{+4.7}_{-5.9}$ &  0.35 $^{+0.14}_{-0.16}$ & 0.27  & $\vline$ & {\bf  NGC 5813}  &  9.8 $^{+3.3}_{-3.8}$ & 0.27 $^{+0.10}_{-0.10}$ & 0.25 \\
NGC 2911      &  2.4 $^{+1.6}_{-1.7}$ &  0.66 $^{+0.29}_{-0.05}$ & 0.20 & $\vline$ & {\bf  NGC 5831}  &  10.7 $^{+3.5}_{-4.6}$ & 0.37 $^{+0.10}_{-0.12}$ & 0.18 \\
{\bf  NGC 2974}      & 14.3 $^{+3.6}_{-1.5}$ &  0.15 $^{+0.06}_{-0.07}$ & 0.25 & $\vline$ & {\bf  NGC 5845}  &  2.5 $^{+0.4}_{-0.5}$ & $>$ $^{+0.05}_{-0.05}$ & 0.23 \\
{\bf  NGC 3091}      &  6.2 $^{+5.2}_{-4.0}$ &  $>$ $^{+0.02}_{-0.01}$ & 0.08  & $\vline$ &  {\bf  NGC 5846}  &  10.2 $^{+3.4}_{-3.7}$ & 0.34 $^{+0.17}_{-0.18}$ & 0.20 \\
NGC 3098      &  3.9 $^{+1.3}_{-1.3}$ &  0.14 $^{+0.08}_{-0.07}$ & 0.05  & $\vline$ & {\bf  NGC 5846A}& 12.3 $^{+4.7}_{-3.8}$ & 0.17 $^{+0.22}_{-0.23}$ & 0.34 \\ 
NGC 3115      &  2.6 $^{+0.0}_{-0.6}$ &  $>$ $^{+0.02}_{-0.06}$ & 0.09 & $\vline$ & NGC 5864  &  5.0 $^{+1.0}_{-1.2}$ &-0.10 $^{+0.07}_{-0.06}$ &-0.13 \\
NGC 3156      &  0.9 $^{+0.6}_{-0.3}$ & -0.17 $^{+0.14}_{-0.10}$ & 0.11 & $\vline$ & {\bf  NGC 5982}  & 12.3 $^{+1.9}_{-2.0}$ & 0.22 $^{+0.04}_{-0.06}$ & 0.19 \\
{\bf  NGC 3193}      &  5.7 $^{+1.7}_{-2.5}$ &  0.48 $^{+0.04}_{-0.13}$ & 0.18  & $\vline$ & {\bf  NGC 6172}  &  1.8 $^{+0.3}_{-1.0}$ & 0.23 $^{+0.13}_{-0.07}$ & 0.01 \\
{\bf  NGC 3226}      & $>$ $^{+3.4}_{-10.1}$ &  -0.02 $^{+0.06}_{-0.07}$ & 0.31 & $\vline$ &{\bf  NGC 6411}  &  8.3 $^{+0.4}_{-0.4}$ & 0.15 $^{+0.05}_{-0.05}$ & 0.11 \\ 
NGC 3245      &  5.8 $^{+2.4}_{-2.7}$ &  0.32 $^{+0.10}_{-0.09}$ & 0.29  & $\vline$ & NGC 7302  &  4.0 $^{+1.0}_{-1.8}$ & 0.33 $^{+0.10}_{-0.10}$ &-0.04 \\
{\bf  NGC 3377}      &  3.6 $^{+0.4}_{-0.5}$ &  0.35 $^{+0.05}_{-0.05}$ & 0.26  & $\vline$ & NGC 7332  &  1.0 $^{+1.0}_{-0.8}$ & $>$ $^{+0.01}_{-0.29}$ &-0.01 \\
{\bf  NGC 3379}      & 10.9 $^{+2.6}_{-2.8}$ &  0.30 $^{+0.07}_{-0.08}$ & 0.21  & $\vline$ & {\bf  NGC 7454}  &  2.8 $^{+1.3}_{-1.2}$ & 0.13 $^{+0.08}_{-0.08}$ & 0.11 \\
NGC 3384      &  6.5 $^{+1.0}_{-1.6}$ &  0.13 $^{+0.05}_{-0.04}$ & 0.21 & $\vline$ & NGC 7585  &  1.3 $^{+0.8}_{-0.6}$ & $>$ $^{+0.14}_{-0.21}$ & 0.07 \\
NGC 3412      &  1.5 $^{+0.5}_{-0.5}$ &  $>$ $^{+0.18}_{-0.10}$ &-0.03 & $\vline$ & {\bf  NGC 7619}  &  5.9 $^{+3.2}_{-2.0}$ & $>$ $^{+0.02}_{-0.13}$ & 0.13 \\
              &                       &                           &      & $\vline$ & {\bf  NGC 7626}  & 13.9 $^{+4.9}_{-2.4}$ & 0.36 $^{+0.06}_{-0.05}$ & 0.29 \\

\hline
\end{tabular}
\vspace{0.2 cm}

\footnotesize{Galaxy names in bold characters are ellipticals. \\
The $>$ symbol represents a lower limit estimate: 15.0 Gyr in Age and 0.67 
dex in [Z/H].}\\
{$\dagger$ Derived assuming model tracks of a constant age of 4 Gyr.}
\end{minipage}
\end{table*}
\normalsize
%%%%%%%%%%%%%%%%%%%%%%%%%%%%%%%%%%%%%%%%%%%%%%%%%%%%%%%%%

%%%%%%%%%%%%%%%%%%%%%%%%%%%%%%%%%%%%%%%%%%%%%%%%%%%%%%%%%%%%%%%%%%%%%%%%%
\subsection{Errors}

It is known that correlated observational errors can masquerade as real trends. Trager \etal (1998) have demonstrated the role of index errors in the determination of ages, metallicities and their correlations with velocity dispersion. Furthermore, the Balmer lines used to infer ages and metallicities in e.g. H$\beta$-[MgFe] diagrams, can be severely affected by emission and its correction is crucial to the determination of galaxy parameters. 
The general effect of correcting H$\beta$ by the presence of emission is to change both the derived ages and metallicities. Age is reduced and [Z/H] increased ([$\alpha$/Fe] is less affected).
It happens so that the information on H$\beta$ errors is commonly used to separate and select the best galaxy data in the literature. The sample used by Trager \etal (1998) had a typical error of $\delta$H$\beta$ = 0.191 \AA.
A reasonable guide is that H$\beta$ must be accurate to $\sim$ 0.1 \AA\ to determine reliable ages and metallicities. 

%%%%%%%%%%%%%%%%%%%%%%%%%%%%%%%%%%%%%%%%%%%%%%%%%%%%%%%%%%%%%%%
%%%%%%%%%%% FIGURE 8: errors!!! %%%%%%%%%%%%%%%%%%%%%%%%%%%%%%%
%%%%%%%%%%%%%%%%%%%%%%%%%%%%%%%%%%%%%%%%%%%%%%%%%%%%%%%%%%%%%%%
\begin{figure*}
\vspace{22.7 cm}
\includegraphics{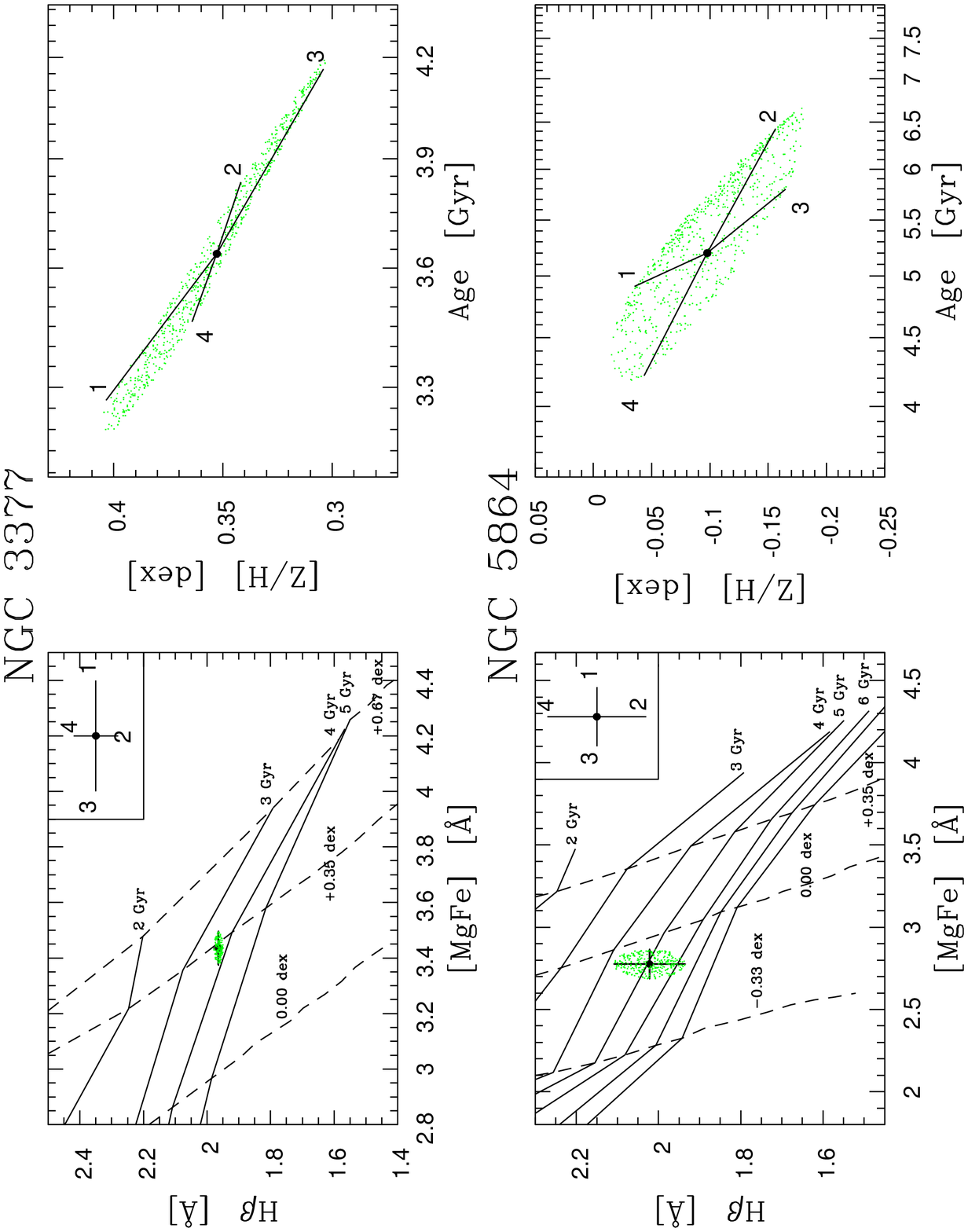}
\includegraphics{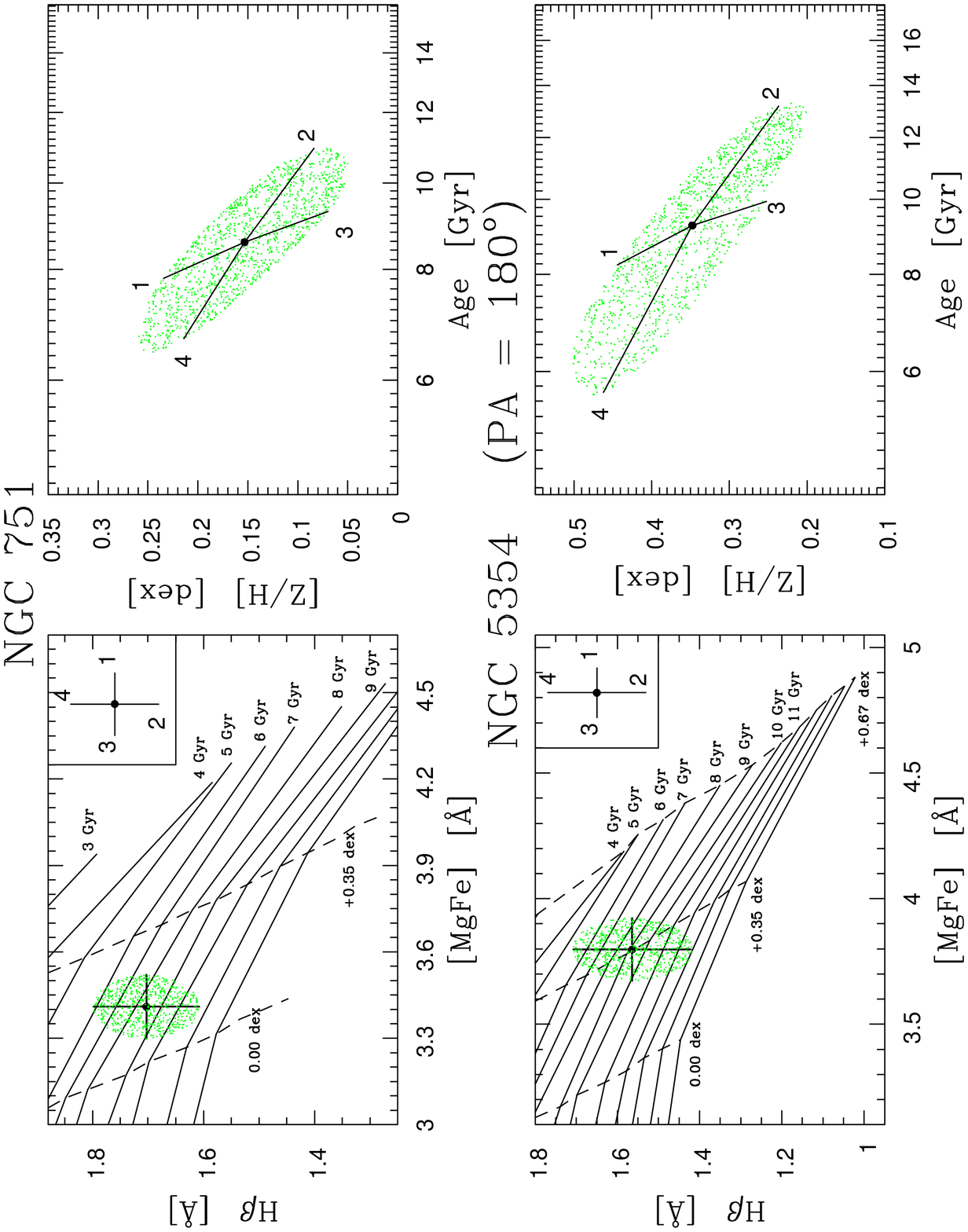}
\caption{Examples of index errors transposed to the age-metallicity diagram. The [MgFe]-H$\beta$ plane is shown on the left-hand side and overplotted are the model grids of TMB03. In each plot we show the position of a galaxy and its error bars in the index-index plane. On the right-hand side panels we show the results of the model grid interpolation to the age-metallicity plane. The 1$\sigma$ error ellipses were obtained from Monte Carlo simulations. }
\end{figure*}
%%%%%%%%%%%%%%%%%%%%%%%%%%%%%%%%%%%%%%%%%%%%%%%%%%%%%%%%%%%%%%%

Our measured observational errors are not as small as Gonz\'alez (1993, $\delta$H$\beta$ = 0.060 \AA) or Kuntschner (1998, $\delta$H$\beta$ = 0.089 \AA), although these are estimated errors. The average H$\beta$-index error in our sample is 0.11 \AA. Our errors have been increased to account for uncertainties in the H$\beta$ emission correction: previous to the emission correction our H$\beta$ mean error was 0.086 \AA. 

It is important to emphasize here that in this work we focus our attention to conclusions drawn from the relative position of galaxies with respect to each other and not on absolute values.

The independent errors in the observables H$\beta$ and [MgFe] from Figure 5 translate into correlated errors in the age {\it vs} metallicity plane. We have calculated the uncertainties in age and metallicity by interpolating new age and [Z/H] values for the points: ({\it H$\beta$$-$error},[MgFe]), ({\it H$\beta$$+$error}, [MgFe]), (H$\beta$, {\it [MgFe]$-$error}) and (H$\beta$, {\it [MgFe]$+$error}). We show in Figure 8 that these four points describe well the uncertainties in the central age and [Z/H] determinations. 

We have performed Monte-Carlo simulations of a large number of random points inside the 1$\sigma$ error contour of a galaxy in the H$\beta$-[MgFe] plane. The random points were then interpolated using the TMB03 model grids; the age and [Z/H] values derived are plotted on the right-hand side panels of Figure 8. Note that the 1$\sigma$ error ellipses on the H$\beta$-[MgFe] plane were transformed into rather twisted ellipses on the age-[Z/H] plane.

%%%%%%%%%%%%%%%%%%%%%%%%%%%%%%%%%%%%%%%%%%%%%%%%%%%%%%%%%%%%%%%
%%%%%%%%%%%% FIGURE 9:  AGE X Z WITH ERRORS %%%%%%%%%%%%%%%%%%%
%%%%%%%%%%%%%%%%%%%%%%%%%%%%%%%%%%%%%%%%%%%%%%%%%%%%%%%%%%%%%%%
\begin{figure*}
\vspace{14.5 cm}
\includegraphics{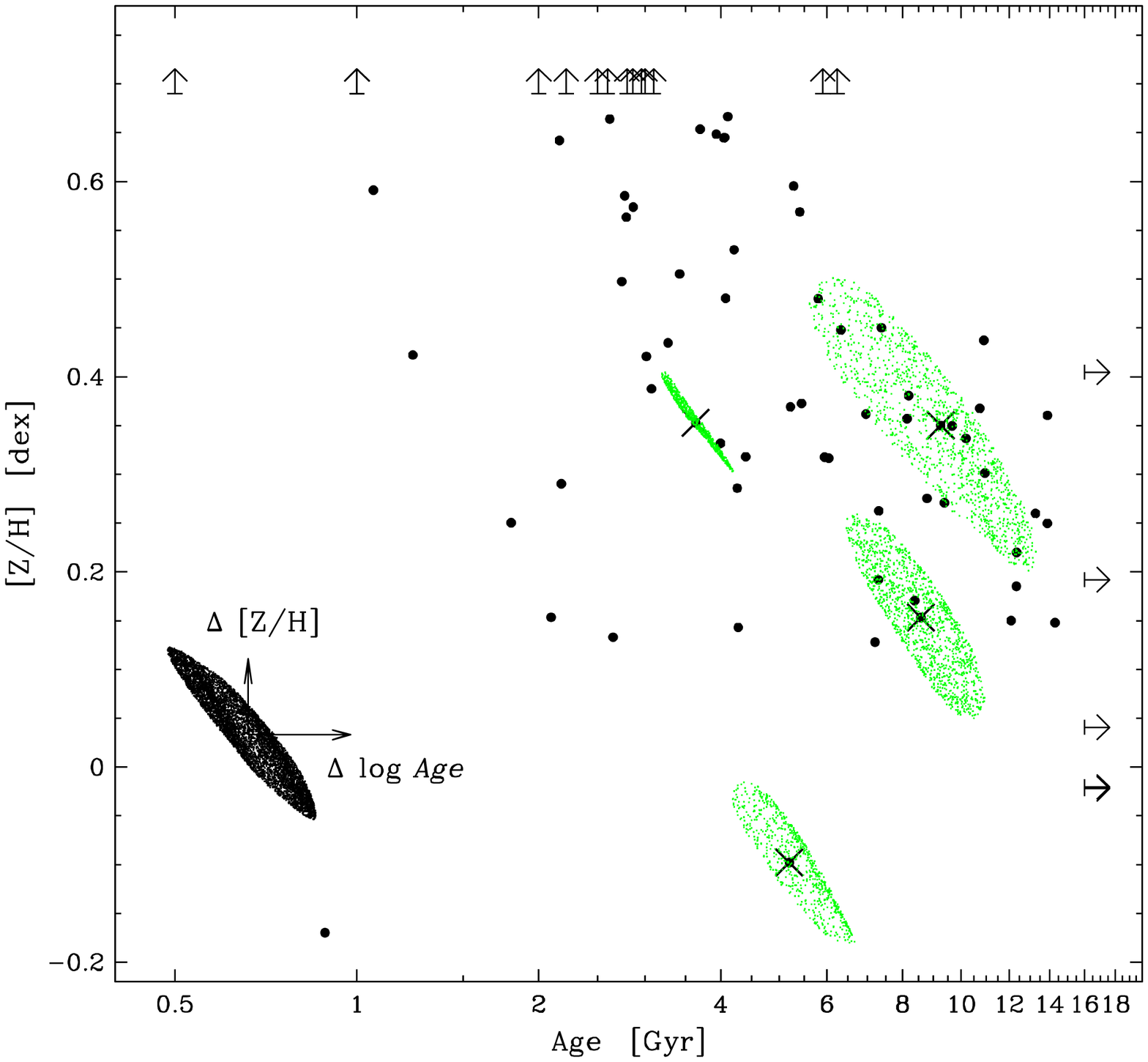}
\caption{Age-metallicity diagram with error ellipses. The error ellipses are the same shown for four galaxies in Figure 8. The slope of the error ellipses is nearly the same as the slope of the overall tendency between age and metallicity (we estimate $\Delta$ log {\it age} $\simeq$ $-$1.57 $\Delta$ [Z/H]), showing that correlated errors can masquerade as real trends.}
\end{figure*}
%%%%%%%%%%%%%%%%%%%%%%%%%%%%%%%%%%%%%%%%%%%%%%%%%%%%%%%%%%%%%%%

The age and [Z/H] values interpolated for ({\it H$\beta$$-$error},[MgFe]), ({\it H$\beta$$+$error}, [MgFe]), (H$\beta$, {\it [MgFe]$-$error}) and (H$\beta$, {\it [MgFe]$+$error}) are also shown in the right-hand side panels. Although these four points do not describe perfectly well the total range of possible age and [Z/H] values inside the 1$\sigma$ error ellipses, they do provide us with a sufficient sampling of the uncertainties involved when transposing errors from the H$\beta$-[MgFe] plane to ages or metallicities.
We have then used the range of values from these four points as the uncertainty in age and [Z/H] determinations presented in Table 2. 

We also show in Figure 9 how the error ellipses of Figure 8 are distributed in the total age-[Z/H] plane. Note that the slope of the error ellipses is very close to the direction of the spread of the points in the  age and metallicity plane (we estimate $\Delta$ log {\it age} $\cong$ $-$1.57 $\Delta$ [Z/H]), indicating that correlated errors can masquerade as real trends.

%%%%%%%%%%%%%%%%%%%%%%%%%%%%%%%%%%%%%%%%%%%%%%%%%%%%%%%%%%%%%%%%%%%%%%%%%
%%%%%%%%%%%%% SECTION 5: ALPHA/FE %%%%%%%%%%%%%%%%%%%%%%%%%%%%%%%%%%%%%%%
%%%%%%%%%%%%%%%%%%%%%%%%%%%%%%%%%%%%%%%%%%%%%%%%%%%%%%%%%%%%%%%%%%%%%%%%%

\section{The [$\alpha$/Fe] ratio}

According to several authors, the central regions of giant E and S0 galaxies show an overabundance of $\alpha$-elements over Fe, indicated in particular by the steep Mg$_2$ index radial gradients (Worthey, Faber \& Gonz\'alez 1992; Trager \etal 2000a; T00b). This finding strongly impacts on the theory of galaxy formation, as super-solar [$\alpha$/Fe] ratios seem to require short star formation time-scales ($\la$ 1 Gyr) that are hardly achieved by current models of hierarchical galaxy formation (e.g., Thomas \& Kauffmann 1999). 

The TMB03 predictions are identical to T00b Model 1, except that the latter 
do not include the $\alpha$-element Ca in the $\alpha$-element group. As pointed out by TMB03, the fractional contribution of Ca to the total metallicity is too small ($\sim$ 0.1 per cent) to change the isochrone and SSP characteristics. 

Super-solar [$\alpha$/Fe] ratios at fixed total metallicity are produced by a decrease of the Fe abundance rather than by an increase of the $\alpha$ 
element abundances ([Z/H] = [Fe/H] + {\it A}[$\alpha$/Fe], where {\it A} $\equiv$ $-$ $\Delta$[Fe/H]/$\Delta$[$\alpha$/Fe]). The actual atomic abundance ratios of the so-called enhanced elements are in fact virtually solar; it is only the 
Fe-peak elements -- which contribute $\sim$ 8\% to the total amount of metals in the Sun -- that are depressed (e.g., Trager \etal 2000a, TMB03).
This result has been interpreted as being due to a fast early chemical enrichment, therefore dominated by Type II supernovae, in the nuclei of giant E and S0 galaxies. Alternative explanations, however, include a changing IMF slope and/or a changing binary fraction for Type I supernova production. Worthey (1998) discusses the possible scenarios thoroughly in his paper, and concludes that not all $\alpha$-elements follow the Mg trend. Worthey (1998) even suggests the existence of a third chemical enrichment source (perhaps another type of SN) other than SN Type I and II, to explain the behaviour of N or of  the V, Sc, Ti trio in the TiO$_{1,2}$ bands. 
Comparing [Mg/Fe] ratios from spiral bulges (McWilliam \& Rich 1994; Jablonka \etal 1996), S0s and elliptical galaxies in the literature (Davies \etal 1993; Carollo \& Danziger 1994; Fisher \etal 1995, 1996), the data available to date seem to indicate that [Mg/Fe] is near zero in most galaxies of all types with velocity dispersion less than $\sim$ 200 km/s (Worthey 1998).

%%%%%%%%%%%%%%%%%%%%%%%%%%%%%%%%%%%%%%%%%%%%%%%%%%%%%%%%%%%%%%%
%%%%%%%%%%% FIGURE 10: Index tracers of $\alpha$-elements %%%%%
%%%%%%%%%%%%%%%%%%%%%%%%%%%%%%%%%%%%%%%%%%%%%%%%%%%%%%%%%%%%%%%
\begin{figure*}
\vspace{20 cm}
\includegraphics{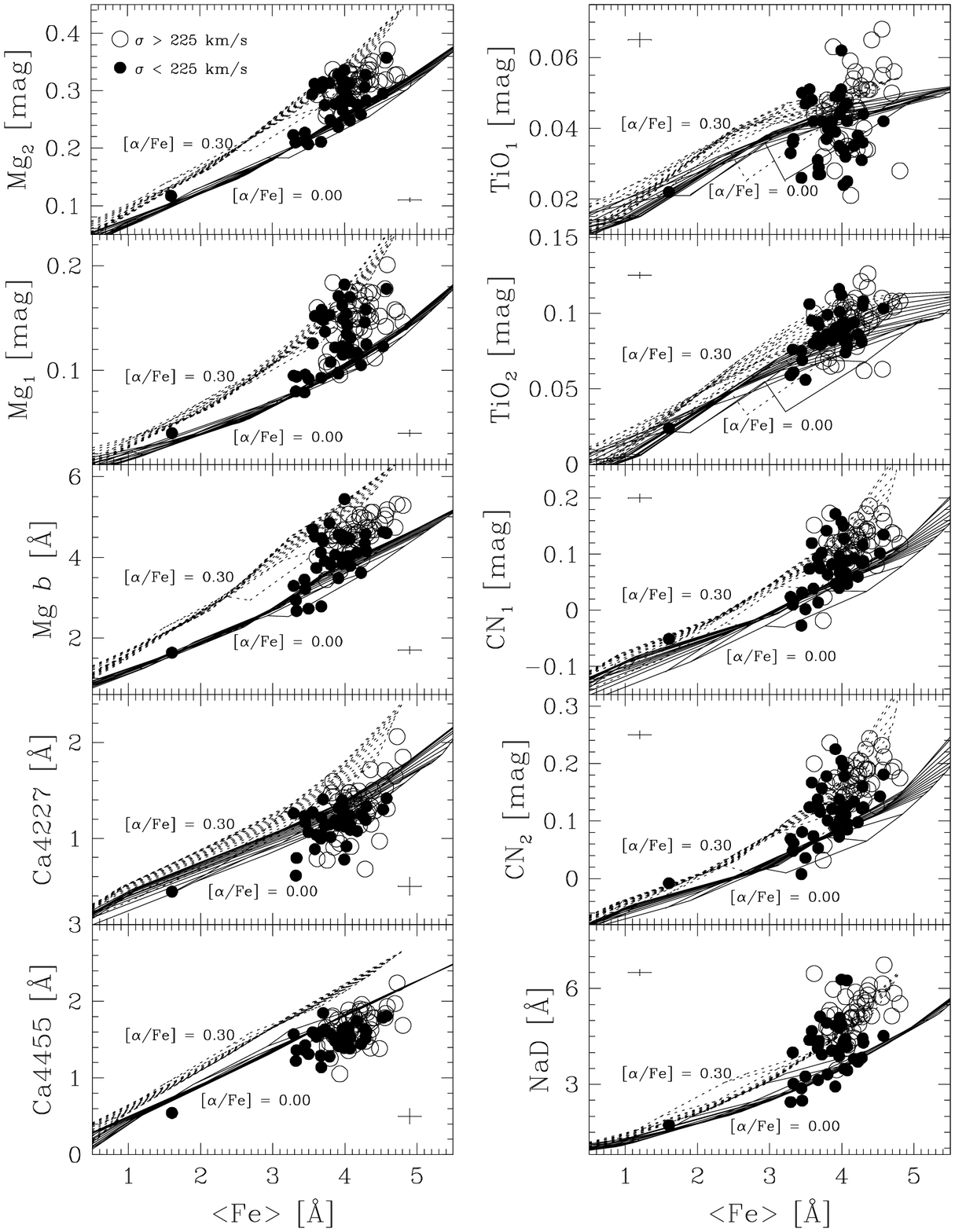}
\caption{Index tracers of $\alpha$-elements {\it vs} $<$Fe$>$. Overplotted are the $\alpha$-enhanced SSPs by TMB03 for [$\alpha$/Fe]=0.0 (solid lines) and [$\alpha$/Fe]=+0.3 (dashed lines). The symbols separate galaxies with $\sigma_0$ smaller or greater than 225 km/s, as indicated in the top-left panel. The outlier galaxy on the bottom-left of all panels is NGC~3156.}
\end{figure*}
%%%%%%%%%%%%%%%%%%%%%%%%%%%%%%%%%%%%%%%%%%%%%%%%%%%%%%%%%%%%%%%

We will examine below the behaviour of some important indices with varying [$\alpha$/Fe] models. In Figure 10 the indices of the $\alpha$-element group are plotted against the average iron index $<$Fe$>$. These diagrams are almost degenerate in age and metallicity, and so are ideal to identify [$\alpha$/Fe] variations by comparing TMB03 models with [$\alpha$/Fe] = 0.0 and +0.3.

Assuming that the difference in central velocity dispersion $\sigma_0$ is determined by the depth of the gravitational potential and ultimately the mass of the galaxy, we separate galaxies according to mass in Figure 10 by assigning different symbols for galaxies with $\sigma_0$ smaller (filled circles) or greater (open circles) than 225 km/s.

The formal [$\alpha$/Fe] ratios per galaxy were determined by interpolating [$\alpha$/Fe] values from the almost degenerate model grids in the Mg$_2$/$<$Fe$>$ and Mg {\it b}/$<$Fe$>$ diagrams, and averaging the resulting values. The interpolation is performed between TMB03 models with [$\alpha$/Fe] = 0.0, +0.3, +0.5 and assuming, for simplicity, only model tracks for an average age of 4 Gyr (i.e., to make the interpolation easier; other model tracks do not alter the results because the Mg/Fe plane is strongly degenerate). We present the derived [$\alpha$/Fe] for the galaxies in our sample in Table 2. It is very difficult to estimate errors for this determination of [$\alpha$/Fe], but examining the errors in Mg and Fe-indices, we assume that on average $\pm$ 0.05 dex in [$\alpha$/Fe] is a very safe upper limit estimate of the uncertainty.

We find that the maximum magnitude of the [Mg/Fe] overabundance is around 0.3 dex, for the central r$_e$/8 extraction. In general, even galaxies with $\sigma_0$ $<$ 225 km/s show Mg enhancement with respect to Fe, and the distinction of both samples (with $\sigma_0$ $<$ 225 km/s and with $\sigma_0$ $>$ 225 km/s) is not always clear in Figure 10. The NaD and CN indices, tracers of Na and mainly N respectively (Tripicco \& Bell 1995, Worthey 1998, TMB03), show the same type of behaviour as Mg with respect to Fe. It appears that Na is more enhanced than Mg, or simply that our measurements are suffering from NaD absorption by interstellar material. 
Except for a few of the less massive galaxies ( $\sigma_0$ $<$ 225 km/s)
the rest show 
$\alpha$-element enhancement (see Figures 10 and 11).
The Ca4227 and Ca4455 indices barely overlay the models for [$\alpha$/Fe] = 0.0, and their different behaviour with respect to the other indices cannot be assigned to [$\alpha$/Fe] enhancement. Both indices are not well calibrated for galaxies, and not convincingly either for globular clusters (Maraston \etal 2003). Ca4227 is sensitive mainly to N, and its Ca-element dependence is yet to be shown. As for Ca4455 -- which is {\it insensitive} to Ca abundance (Tripicco \& Bell 1995), Fe and Cr seem to be the main contributors. The TiO model indices are almost degenerate in the [$\alpha$/Fe] sensitivity, and are badly calibrated even for globular clusters in Puzia \etal (2002). 

In summary, if large [$\alpha$/Fe] ratios require short star
formation time scales for the galaxies, we can say from Figure 10
that, in our sample of mostly low-density environment galaxies,
massive early-type galaxies (high $\sigma_0$) do not obviously
have more rapid star formation (high [$\alpha$/Fe]) than small
(i.e., less massive) early-types, as predicted by hierarchical
galaxy formation models.

%%%%%%%%%%%%%%%%%%%%%%%%%%%%%%%%%%%%%%%%%%%%%%%%%%%%%%%%%%%%%%%
%%%%%%%%%%%% FIGURE 11: CORRELATIONS %%%%%%%%%%%%%%%%%%%%%%%%%%
%%%%%%%%%%%%%%%%%%%%%%%%%%%%%%%%%%%%%%%%%%%%%%%%%%%%%%%%%%%%%%%
\begin{figure*}
\vspace{18 cm}
\includegraphics{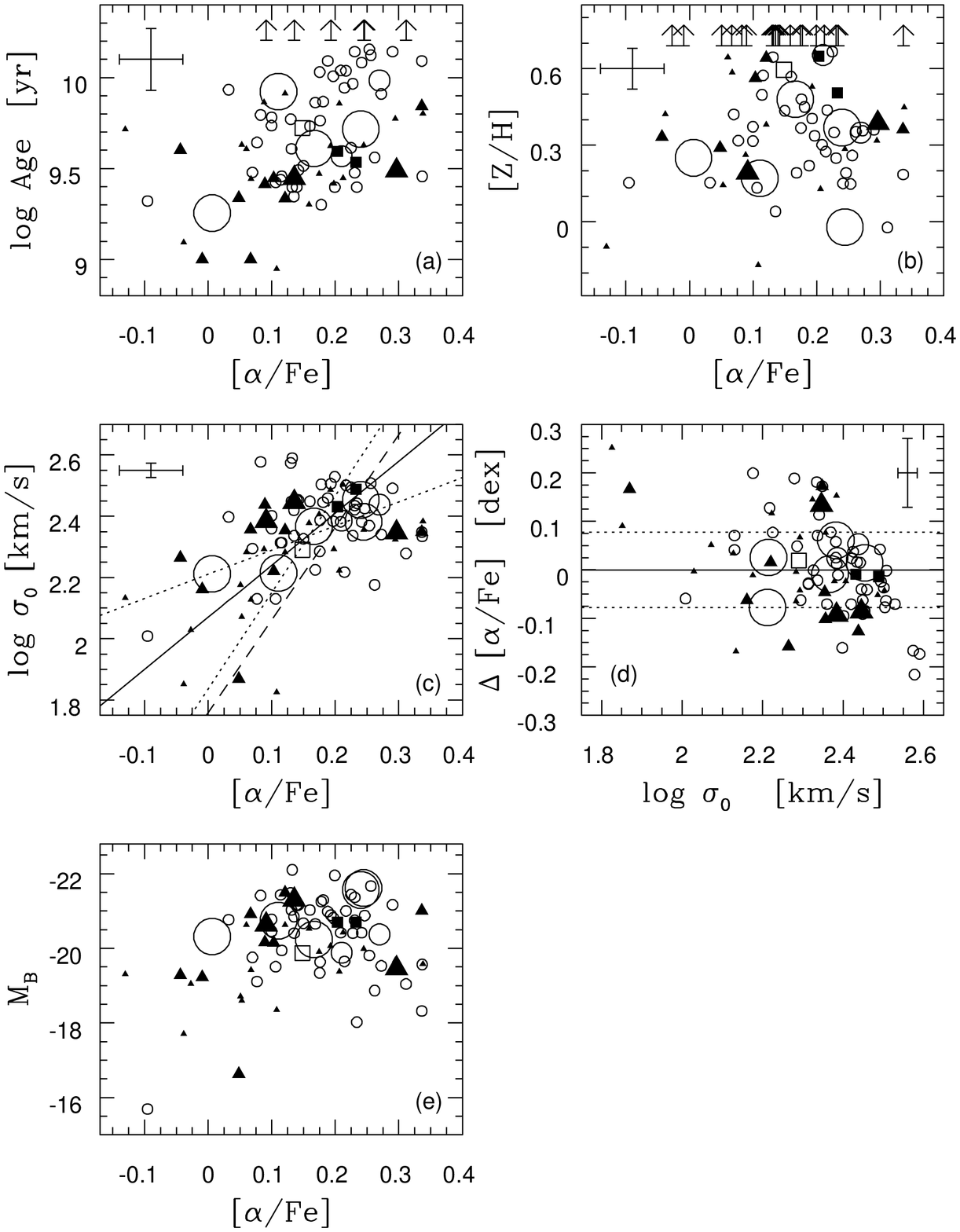}
\caption{Correlations of the enhanced element ratio [$\alpha$/Fe] with (a) age, (b) metallicity, (c) velocity dispersion and (e) absolute magnitude. Panel 
(d) shows the scatter about the solid line fit from panel (c)
and the dotted lines represent the $1\sigma$ scatter 
($\Delta$[$\alpha$/Fe]=0.079 dex. The median [$\alpha$/Fe] ratio of the sample is 0.17 dex.
Circles represent elliptical galaxies and triangles are S0-type galaxies in the sample. Large circles and triangles are isolated galaxies, medium size denote field, and the smallest symbols represent group galaxies. The three Virgo cluster galaxies in the sample, NGC~4365, NGC~4374, NGC~4754, are plotted as square symbols (open square denotes S0, filled square denotes E). The error bars represent the mean uncertainties in log Age (0.17 dex), [Z/H] (0.078 dex), log $\sigma_0$ (0.024 dex) and [$\alpha$/Fe], for which we assume the error estimate of 0.05 dex. }
\end{figure*}
%%%%%%%%%%%%%%%%%%%%%%%%%%%%%%%%%%%%%%%%%%%%%%%%%%%%%%%%%%%%%%%

\

Figure 11 shows the relations of the [$\alpha$/Fe] ratio with age, metallicity, velocity dispersion and magnitude per galaxy. 
There is a trend in that older galaxies are strongly overabundant whereas younger galaxies approach solar abundance ratios (Figure 11-a). A Spearman rank correlation test results in $\rho$ = 0.38 with more than 90\% probability that a correlation is present. It is necessary to note, nevertheless, that part of 
this correlation may be due to the correlated errors discussed before.

The median [$\alpha$/Fe] ratio for the 50 Es is 0.19 $\pm$ 0.08 dex, and for the 30 S0s it is 0.11 $\pm$ 0.11 dex, the uncertainties are the rms values. 

Trager \etal (2000b, T00b), analysing the sample of mainly ``field'' elliptical galaxies from Gonz\'alez (1993), claimed that [$\alpha$/Fe] depends only on $\sigma_0$ and not on age. However, Figure 11-a suggests a SNe II/SNe Ia enhancement ratio higher in the older galaxies of our sample; in other words, there is a hint that older galaxies had short star-formation time scales. On the other hand, there is no clear correlation of [$\alpha$/Fe] with metallicity in Figure 11-b. 

T00b found that [$\alpha$/Fe] and $\sigma_0$ are highly correlated (linear) quantities. Their proposed linear fit to the relation is shown by the dashed line in Figure 11-c.
Examining panel (c) one cannot safely conclude that the previously discussed trend in [$\alpha$/Fe] is as prominent: we find the linear correlation coefficient R = 0.63. Most of the spread in Figure 11-d seems consistent with the errors, thus, for a given $\sigma_0$ most (but not all) galaxies have a similar
 [$\alpha$/Fe] within the errors.
The existence of [$\alpha$/Fe] ratio variations at a given central velocity dispersion 
would imply that the star-formation history of old galaxies at a given $\sigma_0$ is not entirely homogeneous. The [$\alpha$/Fe]-$\sigma_0$ fit to our sample is shown by the solid line in Figure 11-c (slope: 1.7). We have used the bisector (solid line) of the regressions of [$\alpha$/Fe] on $\sigma_0$ and $\sigma_0$ on [$\alpha$/Fe] (represented by the two dotted lines in panel {\it c}). Panel (d) shows the residuals of the [$\alpha$/Fe]-$\sigma_0$ relation (with respect to the solid line of panel {\it c}); the 1$\sigma$ scatter of the relation is $\Delta$ [$\alpha$/Fe] = 0.079 dex, and is represented by the dotted lines in panel (d). 

In Figure 11-e we present the weak relation between [$\alpha$/Fe] and absolute blue magnitude. There is quite a large range of magnitudes for oversolar [$\alpha$/Fe] values. 

In sum these results reinforce the picture of very old and massive galaxies 
having formed in relatively short time scales with small scatter in their
[$\alpha$/Fe] values.

%%%%%%%%%%%%%%%%%%%%%%%%%%%%%%%%%%%%%%%%%%%%%%%%%%%%%%%%%%%%%%%%%%%%%%%%%
%%%%%%%% SECTION 6: HYPERPLANE %%%%%%%%%%%%%%%%%%%%%%%%%%%%%%%%%%%%%%%%%%
%%%%%%%%%%%%%%%%%%%%%%%%%%%%%%%%%%%%%%%%%%%%%%%%%%%%%%%%%%%%%%%%%%%%%%%%%

\section{[Z/H], Age, $\sigma_0$ and [$\alpha$/Fe] hyperplane}

%%%%%%%%%%%%%%%%%%%%%%%%%%%%%%%%%%%%%%%%%%%%%%%%%%%%%%%%%%%%%%%
%%%%%%%%%%%%%%%%%%   FIGURE 12:  [MgFe]     %%%%%%%%%%%%%%%%%%%
%%%%%%%%%%%%%%%%%%%%%%%%%%%%%%%%%%%%%%%%%%%%%%%%%%%%%%%%%%%%%%%
\begin{figure} 
\vspace{12 cm}
\includegraphics{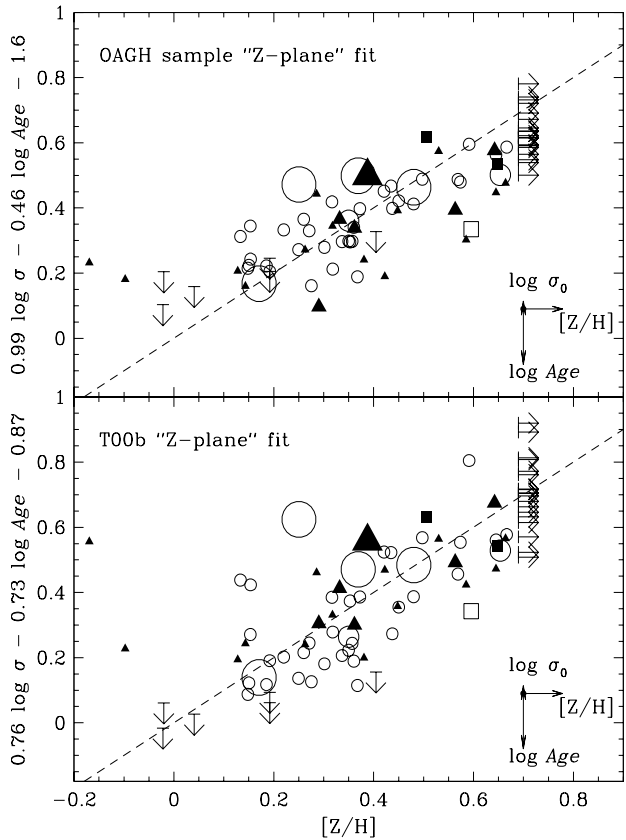} 
\caption{The edge-on view of the metallicity-age-$\sigma$
plane. The dashed line in the top panel is the fit to our data points 
and defines
the plane as [Z/H] = 0.99 log $\sigma_0$ -- 0.46 log {\it Age} --
1.60. The dashed line in the bottom one
is the fit in Trager \etal (2000b). Vectors
of 0.17 in log Age, 0.078 dex in [Z/H] and 0.024 in log $\sigma_0$
represent the average of the overall uncertainties in our
sample. The symbols are the same as in Figure 11. } 
\end{figure}
%%%%%%%%%%%%%%%%%%%%%%%%%%%%%%%%%%%%%%%%%%%%%%%%%%%%%%%%%%%%%%%

In a very interesting paper, Bressan \etal (1996) explored the four dimensional space which involved H$\beta$, [MgFe], (1550-V) and log $\sigma_0$ in elliptical galaxies. Forbes \& Ponman (1999) were among the first to explore the relation between $\sigma_0$ and stellar population age derived from model comparison. Later, T00b performed a principal component analysis (PCA; other references to PCA: Whitney 1983, Ronen \etal 1999) on four variables, namely log $\sigma_0$, log {\it Age}, [Fe/H] and [E/Fe]; i.e. velocity dispersion, age, metallicity and element-enhanced ratio which they called E/Fe (here we adopt the notation $\alpha$/Fe). T00b found that most of the variance (91 percent) in their sample of local ellipticals can be explained with just two parameters, i.e., the sample is basically bi-parametric in the four-dimensional space $\sigma_0$, {\it Age}, [Fe/H] and [E/Fe], in other words, the galaxies are located in a hyperplane in this four-dimensional space that they named, the metallicity hyperplane.

Given the importance of this finding, we have performed a PCA in our sample data. The results are presented in Table 3 and Figure 12.

%%%%%%%%%%%%%%%%%%%%%%%%%%%%%%%%%%%%%%%%%%%%%%%%%%%%%%%%%%%%%%%%
%%%%%%%%%%%%%%%% TABLE 3: PCA %%%%%%%%%%%%%%%%%%%%%%%%%%%%%%%%%%
%%%%%%%%%%%%%%%%%%%%%%%%%%%%%%%%%%%%%%%%%%%%%%%%%%%%%%%%%%%%%%%%
\begin{table}
\scriptsize
\centering
\small
\baselineskip=64pt
\begin{minipage}{90mm}
\caption{Principal component analysis: the hyperplane.}
\vskip 0.2cm
\begin{tabular}{@{}lrrrr@{}}
\hline
\multicolumn{1}{l}{} &
\multicolumn{1}{c}{log $\sigma_0$} &
\multicolumn{1}{c}{log Age} &
\multicolumn{1}{c}{[Z/H]} &
\multicolumn{1}{c}{[$\alpha$/Fe]} \\
\hline  
Covariance matrix:  &            &           &         &      \\    
log $\sigma_0$    &    1.0000 &   0.5944  &  0.4290 &   0.5757 \\
log Age           &    0.5944 &   1.0000  & -0.1919 &   0.5097 \\
$[Z/H]$         &    0.4290 &  -0.1919  &  1.0000 &   0.1709 \\
$[\alpha/Fe]$ &    0.5757 &   0.5097  &  0.1709 &   1.0000 \\
\hline
PCA results:  &            &           &         &          \\
              &     PC1    &    PC2    &   PC3   &     PC4  \\
\hline
log $\sigma_0$    &    0.6135  &   0.1700  &  0.3938 &   0.6630 \\
log Age           &    0.5154  & -0.4987   &  0.3875 &  -0.5792 \\
$[Z/H]$         &    0.2206  &  0.8477   &  0.0887 &  -0.4742 \\
$[\alpha/Fe]$ &    0.5561  & -0.0616   &  -0.8288&   -0.0065\\
\hline
Eigenvalue  & 2.139  &   1.1668 &   0.4621 &   0.1654 \\
Variance    & 0.5438 &   0.2966 &   0.1175 &   0.0421 \\
Cumulative  & 0.5438 &   0.8405 &   0.9579 &   1.0000 \\
\hline
\end{tabular}
\vspace{0.2cm}
\end{minipage}
\end{table}
%%%%%%%%%%%%%%%%%%%%%%%%%%%%%%%%%%%%%%%%%%%%%%%%%%%%%%%%%%%%%%%%

It is interesting to note from an inspection of the covariance matrix that while log $\sigma$ is significantly correlated to the other three variables (log Age, [Z/H] and [$\alpha$/Fe]), both log Age and [$\alpha$/Fe] are not correlated to [Z/H] in our sample.
A similar conclusion regarding the behaviour of log $\sigma_0$, [Z/H] and log Age can be obtained from Figure 13 where the panels show the sample segregated according to $\sigma_0$.

The least square fit to our hyperplane is:

\begin{equation} 
[Z/H] = 0.99(\pm0.15) log \sigma_0 - 0.46(\pm0.08) log {\it Age} - 1.60(\pm0.40)
\end{equation}

%%%%%%%%%%%%%%%%%%%%%%%%%%%%%%%%%%%%%%%%%%%%%%%%%%%%%%%%%%%%%%%%%%%%%%
%%%%%%%%%% FIGURE 13: Age-metallicity plane for constant SIGMA %%%%%%%
%%%%%%%%%%%%%%%%%%%%%%%%%%%%%%%%%%%%%%%%%%%%%%%%%%%%%%%%%%%%%%%%%%%%%%
\begin{figure*} 
\vspace{6 cm}
\includegraphics{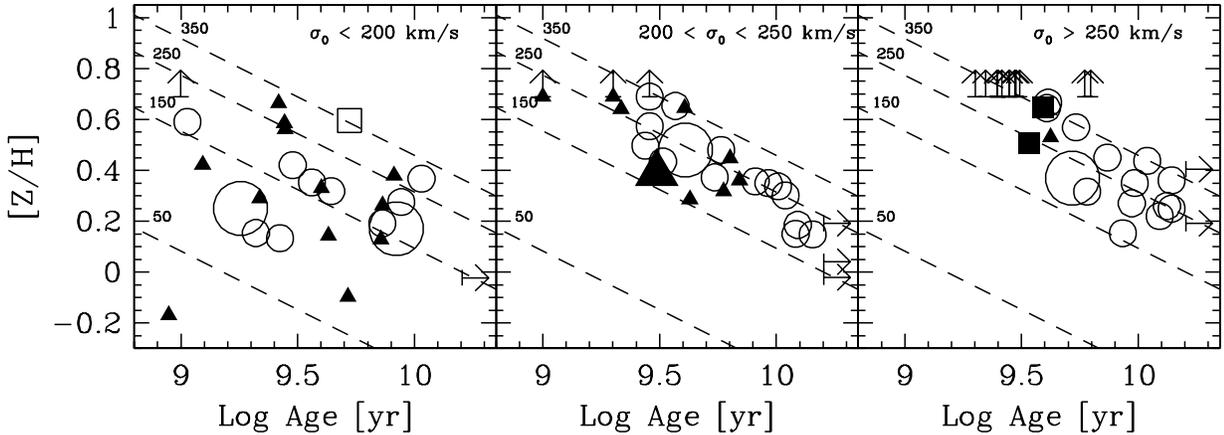}
\caption{Age-metallicity
plane for constant $\sigma_0$. Circles represent elliptical galaxies and
triangles are S0-type galaxies in the sample. Large circles and triangles
are isolated galaxies, medium size denote field, and the smallest symbols
represent group galaxies. The three Virgo cluster galaxies in the sample,
NGC~4365, NGC~4374, NGC~4754, are plotted as square symbols. The
panels separate the galaxies in bins of $\sigma_0$; the dashed lines are
fits from eq. 1, assuming constant $\sigma_0$ values as labelled.} 
\end{figure*}
%%%%%%%%%%%%%%%%%%%%%%%%%%%%%%%%%%%%%%%%%%%%%%%%%%%%%%%%%%%%%%%

In linear terms, this equation can be translated to the approximate law:

\begin{equation} 
{Z \propto \sigma_0 \cdot {\it Age}^{-0.5} } \\
\end{equation}
 
If we assume constant values for $\sigma_0$ and plot eq. 1 in
the Age-metallicity plane, we find the interesting relations shown in
Figure 13. The existence of the Z-plane indicates that there exists an
age--metallicity relation {\it for each value of $\sigma_0$}. This relation
implies, as shown in Figure 13, that at a fixed $\sigma_0$,
metal-rich galaxies have a  younger stellar population in their central
regions. T00b have found the projection of the hyperplane to be  [Z/H] = 0.76($\pm$0.13) log$\sigma_0$ - 0.73($\pm$0.06) log {\it Age} - 0.87($\pm$0.30), where {\it Age} is always taken in Gyr.

From the correlation matrix, the strongest effect seems to be a change in 
log age (and not in [Z/H]) in the sense that age increases going from small $\sigma_0$ ellipticals to large $\sigma_0$ ones. Even though there are several 
lower limits in the right panel of Figure 13 (massive galaxies that fall 
outside the model grids) the effect is visible between the left and central panels.

%%%%%%%%%%%%%%%%%%%%%%%%%%%%%%%%%%%%%%%%%%%%%%%%%%%%%%%%%%%%%%%
%%%%%%%%% FIGURE 14: PCA %%%%%%%%%%%%%%%%%%%%%%%%%%%%%%%%%%%%%%
%%%%%%%%%%%%%%%%%%%%%%%%%%%%%%%%%%%%%%%%%%%%%%%%%%%%%%%%%%%%%%%
\begin{figure*} 
\vspace{16 cm}
\includegraphics{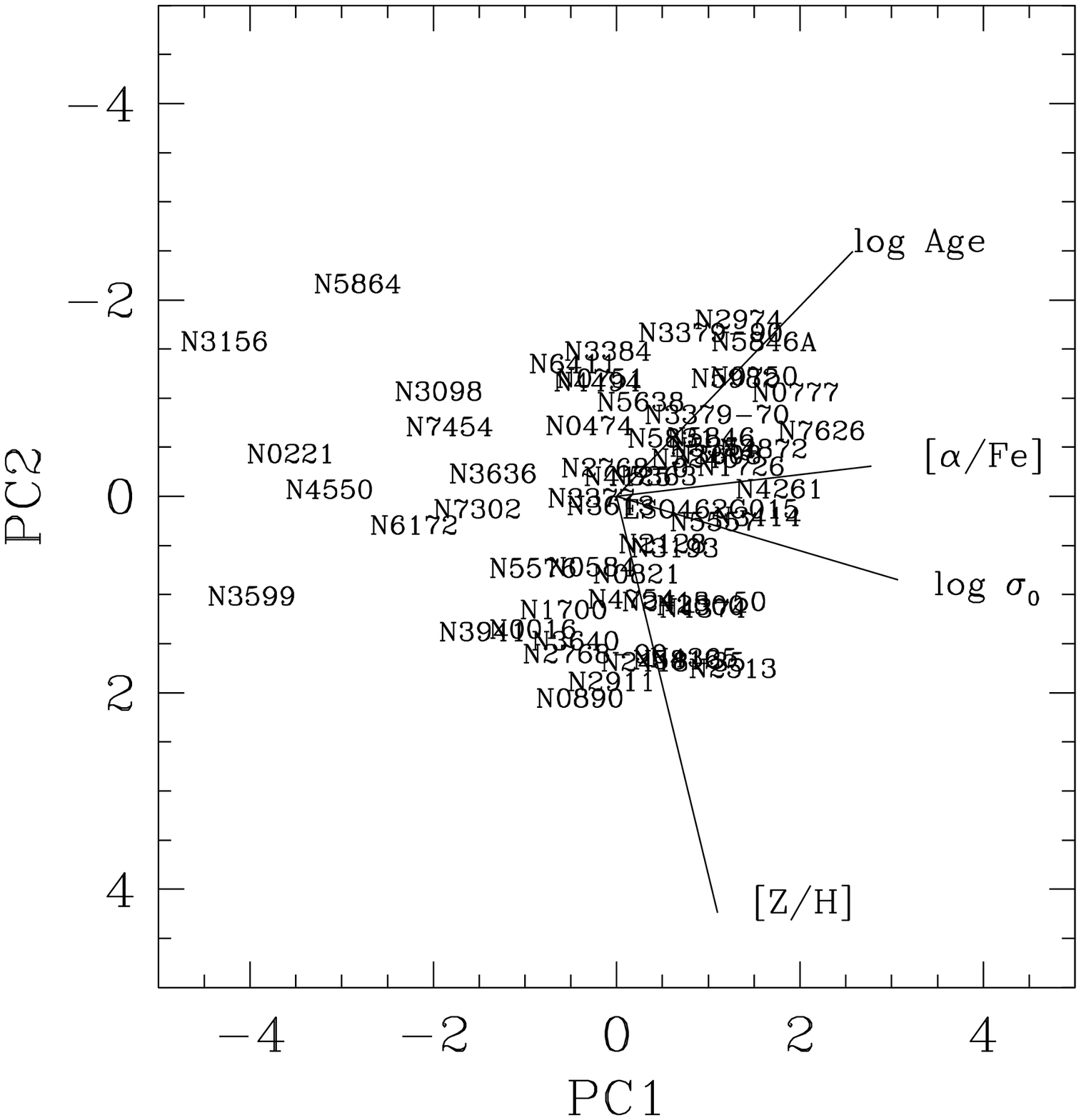}
\includegraphics{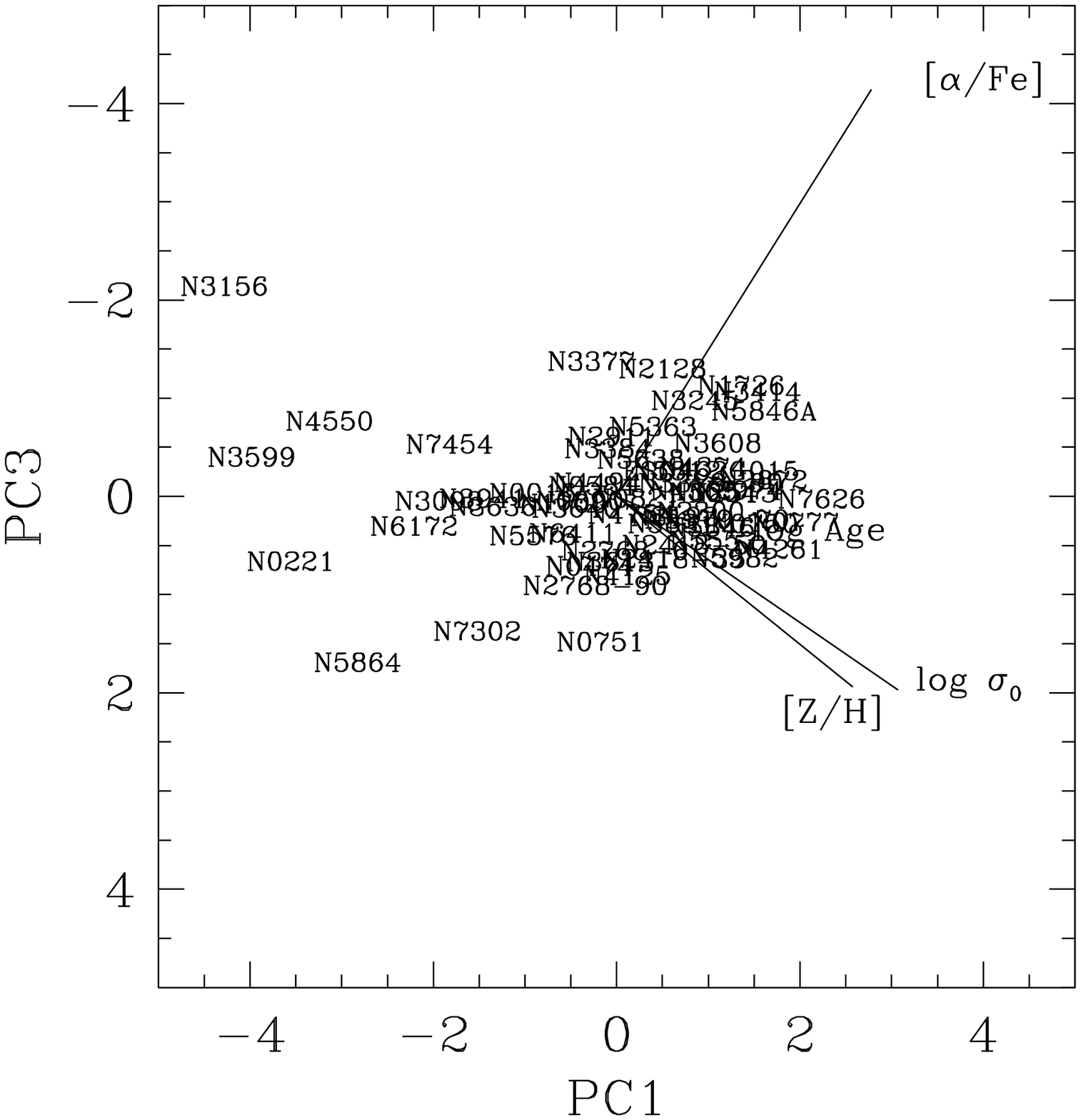}
\includegraphics{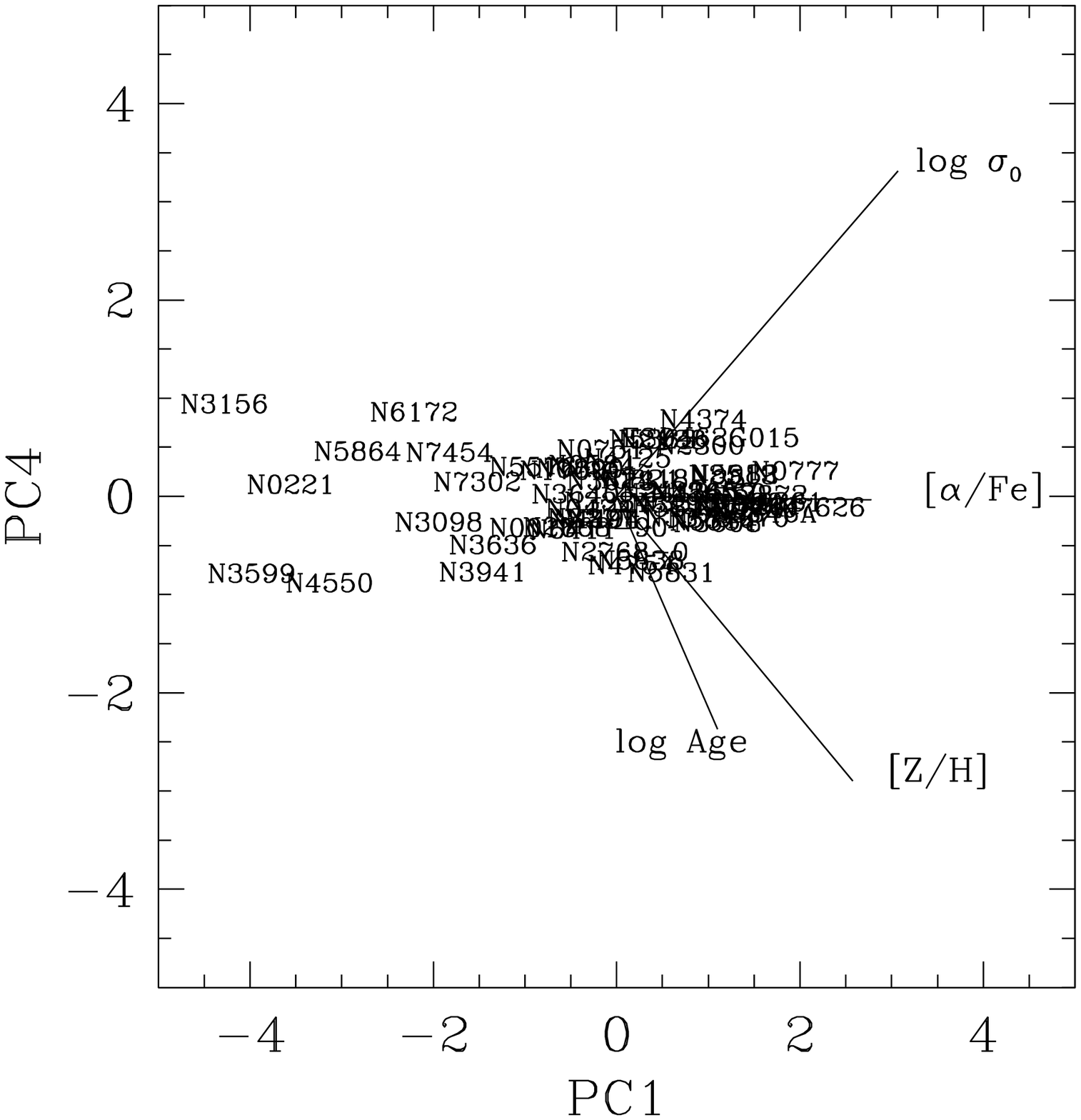}
\caption{Principal Component Analysis results on the OAGH sample. The position of the galaxies in the diagrams is denoted by their names. The top panels represent the face-on (right) and edge-on (left) projections of the log {\it Age}-[Z/H]-log $\sigma$-[$\alpha$/Fe] hyperplane. Projections of the four basic variables are shown as arrows in the direction of increase (for log {\it Age}, this arrow points in the direction of older galaxies). Velocity dispersion and enhancement ratio dominate the first principal component (PC1), while age and metallicity dominate the second (PC2). The third (PC3) and fourth (PC4) principal components contribute less than 12\%  to the overall variance in the log{\it Age}-[Z/H]-log $\sigma$-[$\alpha$/Fe] space.} 
\end{figure*}
%%%%%%%%%%%%%%%%%%%%%%%%%%%%%%%%%%%%%%%%%%%%%%%%%%%%%%%%%%%%%%%

This is an interesting result, and cannot be a reflection of the well known Mg-$\sigma$ relation. If that was the case one would expect a stronger correlation between $\sigma_0$ and metallicity, given that Mg is the dominant index in the [Z/H] estimate in this work using TMB03 models.

Nevertheless, one must be aware in dealing with projections of hyperspaces that the apparent correlation may only be the product of the distribution of points in the hyperspace projected onto a particular plane.

To answer this point we performed a Principal Components Analysis, the results
of which are given in Table 3. The
first two principal components contain 84 percent of the total variance
indicating that the ellipticals in our sample are distributed in a
flattened volume in the 4-D space. This is further illustrated
in Figure 14 where we show the face-on and edge-on
views of the projected variables. To simplify comparisons with T00b
we have used similar orientation in our diagrams. It is interesting how
similar are our results and those of T00b, strengthening the reality of the metallicity hyperplane.

\

Summarising, metallicity is a linear function of both log {\it Age} and log $\sigma_0$, i.e.,  Z $\propto$ $\sigma_0$$\cdot${\it Age}$^{-0.5}$, which defines the so called ``Z-plane''. Following the fit we derived for our sample, the contours of constant $\sigma_0$ have the slope $\Delta$log{\it Age}$\approx$$-$2.2$\Delta$[Z/H]. This is close to the ``3/2 relation'' of Worthey (1994), which expresses
trajectories in the log {\it Age} - [Z/H] space along which colours and
line-strengths remain roughly constant. Thus, the Z-plane correlation
predicts that line-strengths should be constant along trajectories of
constant $\sigma_0$ in the Z-plane.

The enhancement ratio, [$\alpha$/Fe], is a linear function of log $\sigma_0$, increasing towards high-$\sigma_0$ galaxies, and should be also responsible for some of the scatter in Figure 12.

%%%%%%%%%%%%%%%%%%%%%%%%%%%%%%%%%%%%%%%%%%%%%%%%%%%%%%%%%%%%%%%%%%%%%%%%%
%%%%%%%% SECTION 7: DISCUSSION %%%%%%%%%%%%%%%%%%%%%%%%%%%%%%%%%%%%%%%%%%
%%%%%%%%%%%%%%%%%%%%%%%%%%%%%%%%%%%%%%%%%%%%%%%%%%%%%%%%%%%%%%%%%%%%%%%%%

\section{Discussion}

Kuntschner (1998) found that the Fornax  cluster elliptical galaxies appear to be roughly coeval with some evidence for slightly younger ages for the most metal-rich ones. Using Lick/IDS indices and Worthey (1994) SSP models, he estimated the ages to be around 8 Gyr, and the galaxies showed a sequence of metallicities from -0.1 to +0.35 in [Fe/H]. This result is consistent with the conventional view of old, coeval elliptical galaxies where the metallicity scales roughly with the size of the galaxy. However, the lenticular galaxies in Fornax have {\it luminosity weighted} ages that are smaller than those of the ellipticals, spanning from less than $\sim$ 2 Gyr to 8 Gyr, and also covering a large range of metallicities from -0.5 $<$ [Fe/H] $<$ +0.5.
For the purpose of comparison with literature data, Kuntschner (1998) also observed the galaxy NGC~3379. The age of NGC~3379 in Kuntschner's analysis is $\sim$ 8.4 Gyr, whereas we have estimated $\sim$ 10.0 Gyr (with TMB03 models). 
We feel that comparing age estimates from different authors (which may have used different models) is not a reliable procedure; to check for systematic differences between cluster and field environment we prefer to compare the distribution of the galaxies in the [MgFe]-H$\beta$ plane with the Fornax sample distribution, as both are in the Lick/IDS index system. 

%%%%%%%%%%%%%%%%%%%%%%%%%%%%%%%%%%%%%%%%%%%%%%%%%%%%%%%%%%%%%%%
%%%%%%%%%%%%%%%%% FIGURE 15: COMPARISON %%%%%%%%%%%%%%%%%%%%%%%
%%%%%%%%%%%%%%%%%%%%%%%%%%%%%%%%%%%%%%%%%%%%%%%%%%%%%%%%%%%%%%%
\begin{figure*}
\vspace{12. cm}
\includegraphics{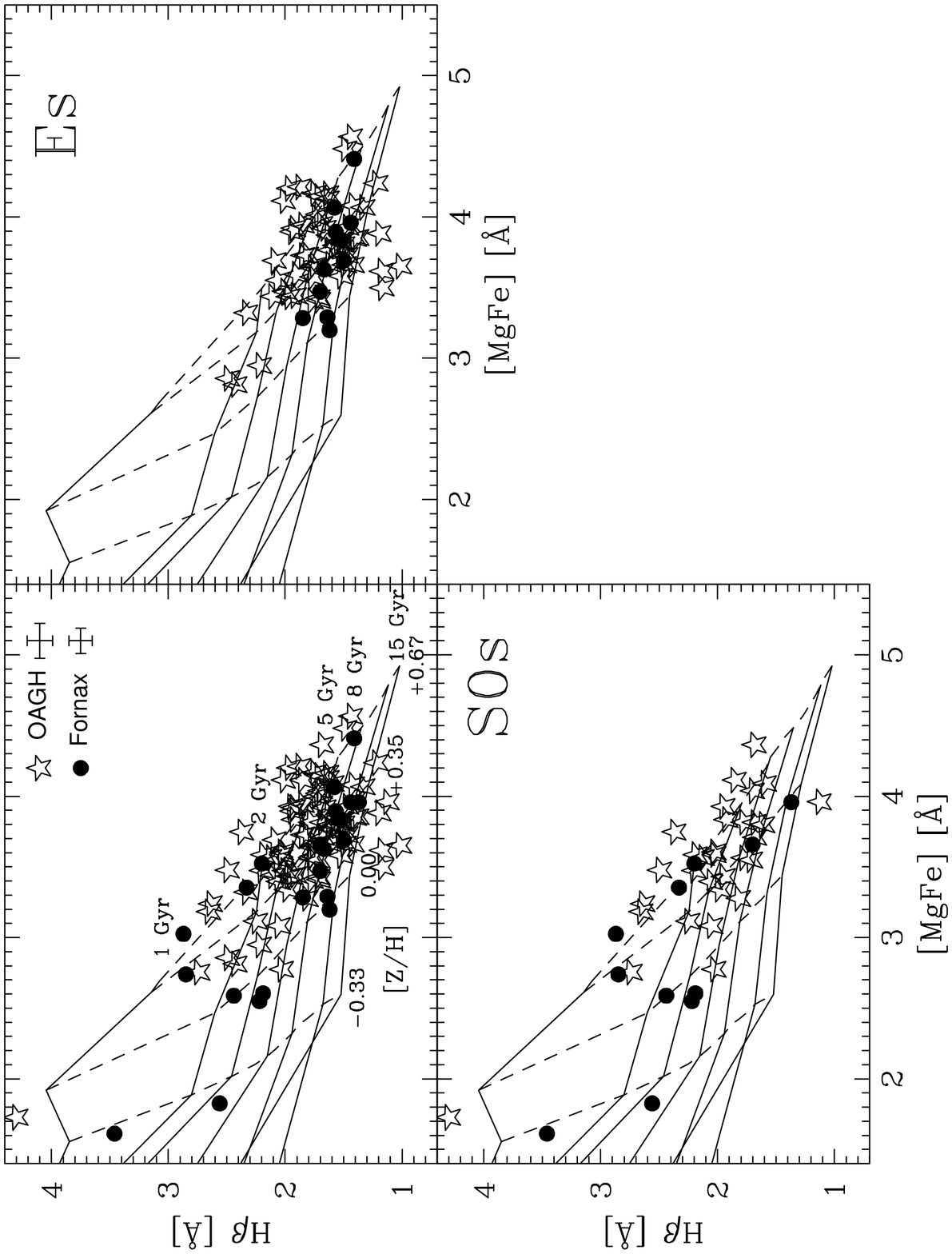}
\caption[Age and metallicity distributions of Fornax and OAGH samples.]
{Comparison between the distribution in the H$\beta$-[MgFe] plane of Fornax and OAGH samples. The model grids are from TMB03, with [$\alpha$/Fe] = 0.0, in age steps of 1, 2, 3, 5, 8, 12, 15 Gyr and metallicity steps of -0.33, 0.00, +0.35 and +0.67 dex. The bottom panel shows only S0 galaxies in both samples, and the right-hand panel shows only elliptical galaxies.}
\end{figure*}
%%%%%%%%%%%%%%%%%%%%%%%%%%%%%%%%%%%%%%%%%%%%%%%%%%%%%%%%%%%%%%%

In Figure 15 we show a comparison between H$\beta$ and [MgFe] indices for the Fornax and OAGH samples. The bottom panel of Figure 15  shows only S0 galaxies in both samples, and the right-hand panel shows only elliptical galaxies.

We estimate that the average age for our sample of group, field and isolated galaxies (5.8 Gyr) is younger than Fornax cluster ellipticals ($\sim$ 9-10 Gyr). Secondly, both Es and S0s in our sample cover a large range of ages {\it and} metallicities. Although our galaxies span wider ranges in age and [Z/H] than the Fornax galaxies, the Es and S0s in the OAGH sample appear to have on average more similar stellar populations than the very distinct behaviour of Es and S0s in Fornax.

In Section 4 we have estimated that S0s present a younger {\it mean} age than Es. It is important to emphasize here that the line-strength indices reflect only the integrated, luminosity weighted properties in a galaxy. As young populations are more luminous per unit mass than old ones, a {\it small} (in mass) young population can dramatically change the strength of indices, in particular H$\beta$. de Jong \& Davies (1997) and Trager (1997) have demonstrated this and also showed that the strong H$\beta$ galaxies in Gonz\'alez (1993) sample tend to have disky isophotes. de Jong \& Davies (1997) suggested that ongoing star-formation might be associated with the presence of a disk. In the OAGH sample, elliptical galaxies that present young ages and hints of a stellar disk are NGC~315 (Peletier \etal 1990; our age: 2.5 Gyr), NGC~584 (Michard \& Marchal 1994; 3.8 Gyr), NGC~720 (Goudfrooij \etal 1994; 3.0 Gyr) , NGC~821 (Ravindranath \etal 2001; 4.0 Gyr), NGC~3377 (Scorza 1993; 3.6 Gyr), NGC~3610 (Scorza \& Bender 1990; 1.8 Gyr), NGC~3613 (Goudfrooij \etal 1994; 5.6 Gyr), NGC~3640 (Goudfrooij \etal 1994; 2.5 Gyr), NGC~4365 (Forbes \etal 1995; 3.6 Gyr), NGC~4374 (Zirbel \& Baum 1998; 3.8 Gyr), NGC~5322 (Goudfrooij \etal 1994; 2.4 Gyr), NGC~5845 (Reid \etal 1994; 2.5 Gyr), i.e., 60\% of elliptical galaxies with ages younger than 4 Gyr show photometric evidence of a stellar disk component in their morphologies.

NGC~3156 has a very distinct behaviour with respect to the bulk of the sample. As we have noted in some plots, this is the youngest and most metal-poor elliptical galaxy in the sample (NGC~5864 shows equivalently  low metallicity but with larger error bars, and substantially older age). According to the calculations of Poggianti \& Barbaro (1997) {\it strong} Balmer absorption is created mainly at the main sequence turn off and cannot originate from other evolutionary stages (at least not for relatively young systems). Note that we are referring to {\it strong} Balmer absorption in a galaxy with a {\it young} stellar population. In many cases, Horizontal Branch (HB) stars can affect Balmer lines rather severely (see Maraston \etal 2003 for TMB03 model calibration of the Balmer lines). However, in the case of NGC~3156 it would be necessary to have a significant population of HB stars to produce such a young age. In fact, Maraston \etal (2003) have determined rather old ages ($\sim$ 8 Gyr) for two globular clusters NGC~6441 and NGC~6388 which are known to exhibit an important tail of warm HB stars. Furthermore, Burstein \etal (1988) have shown that the ultraviolet excess in early-type galaxies (which is now known to be associated with HB stars; O'Connell \etal 1999) correlates well with Mg$_2$ (NGC~3156 shows weak Mg$_2$ absorption [cf. Paper I], and poor metallicity).
 Hence we believe that the young age estimated from the model predictions  for NGC~3156 is reliable. This is a disk dominated S0 galaxy, with a dusty and faint bulge, located in a group and shows no obvious interaction features. It is a galaxy that shows a high V/$\sigma_0$ value (=1.2; Paper I).

%%%%%%%%%%%%%%%%%%%%%%%%%%%%%%%%%%%%%%%%%%%%%%%%%%%%%%%%%%%%%%%
%%%%%%%%%%%%%%%   FIGURE 16:  sigma age Z   %%%%%%%%%%%%%%%%%%%
%%%%%%%%%%%%%%%%%%%%%%%%%%%%%%%%%%%%%%%%%%%%%%%%%%%%%%%%%%%%%%%
\begin{figure} 
\vspace{8.5 cm}
\includegraphics{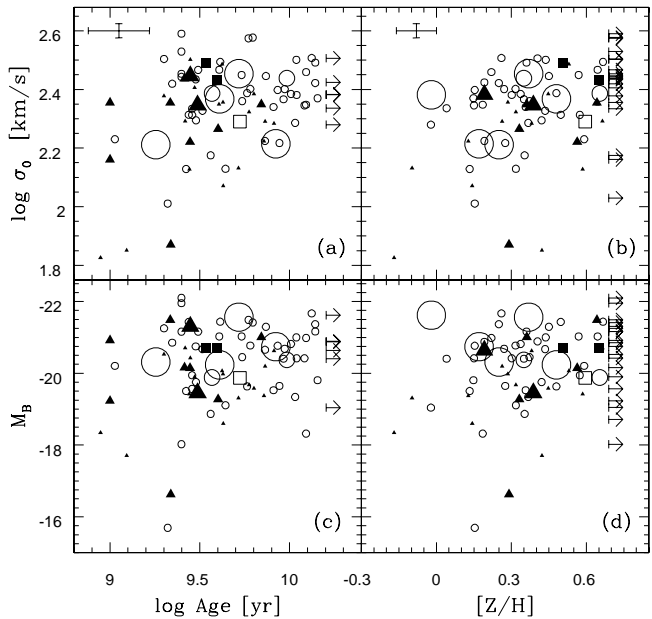} 
\caption{Panels {\it a} and {\it b} present the central velocity dispersion against Age and metallicity determinations for the r$_e$/8 aperture extraction. Panels {\it c} and {\it d} show the absolute blue magnitude against Age and [Z/H].  
The symbols are the same as in Fig.11.} 
\end{figure}
%%%%%%%%%%%%%%%%%%%%%%%%%%%%%%%%%%%%%%%%%%%%%%%%%%%%%%%%%%%%%%%

The relations between Age and metallicity [Z/H]
against velocity dispersion and absolute blue magnitude 
(calculated from apparent magnitudes and our
determinations of radial velocity given in Paper I, and assuming
H$_o$ = 75 km/s/Mpc) are shown in
Figure 16. We observe a trend in that the oldest and more metal rich
galaxies tend to have higher velocity dispersions.
Caldwell, Rose \& Concannon (2003) and Trager \etal (2000a,b) claim 
that there is a trend for lower $\sigma_0$ (or less massive elliptical)
galaxies to have younger ages, contrary to what is found by 
Kuntschner (2000) and TF02 with their samples. Caldwell \etal (2003)
also note a modest (if any) trend in [Z/H] with $\sigma$.
Ours and these previous results, however, are not conclusive enough
to place key constraints on age distributions
predicted by numerical simulations of galaxy formation, but we refer
the reader to a discussion of this point in TF02.

In Section 5 we have seen that there is a marginal tendency for
massive galaxies to show larger [$\alpha$/Fe] values than less massive
ones.
Worthey (1992,1998) detail several processes that can produce  an  
$\alpha$-element enhancement as a function of galaxy mass.
They are mostly related to the different contributions towards the
chemical enrichment of galaxies by supernovae types Ia and II.
SN~Ia (with progenitors less  massive than 5-10 M$_{\odot}$) release
Fe-peak elements after a time delay of about a few times 10$^8$ years.
Type II  supernovae, on the other hand, originate from more massive stars
and contribute earlier to the abundance of $\alpha$-elements and Fe.
The ratio of SNe type II to type Ia, which can vary from massive to less
massive galaxies, somehow determines the [$\alpha$/Fe] abundances.
There are several scenarios proposed for the variation of type~II/type~Ia
ratio (e.g.~Arimoto \& Yoshii 1987) including winds, amount of binaries, even
IMF variations due to environmental causes. Different time-scales
for the formation of SNe Types~II and Ia, allow the prediction of a 
correlation of Mg strength with dynamical crossing time (Worthey 1998)
or, if including the effects of winds (Arimoto \& Yoshii 1987) with the
potential well of the galaxy.
An environmentally controlled IMF predicts good correlation with 
$\sigma$ (Paper I).
While  Mg$_2$  versus $\sigma$ is a well known tight  correlation,
$<$Fe$>$ versus $\sigma$ is weaker. This suggests that the basic relation
is between the Mg abundance and the velocity dispersion.
Contrary to Worthey (1998), who finds that [Mg/Fe] vs.~$\sigma$ does not 
depend on 
galaxy type, supporting the statement above, we find that ellipticals
have   [$\alpha$/Fe] ratios slightly higher than S0s, on average.

Other indices that correlate well with  $\sigma$ are  NaD and CN (Worthey 
1998). Matteucci \& Padovani (1993) claim that a change in the IMF slope
 of 0.4 could produce   a 0.3 dex change in $\alpha$-element abundances.
It remains to be proved that such a change in the IMF slope could be 
consistent with observations.

The metallicity of the interstellar medium can provide a constraint to decide
between the time formation difference hypothesis and the IMF variation one
(Worthey 1998). We have to resort to X-ray abundance determinations where 
we encounter difficulties with modelling the Fe atom (Loewenstein \& 
Mushotzky 1996;  Arimoto \etal 1997) although progress has been achieved in 
that direction (Matsushita \etal 2000).

Finally, in the previous Sections we have seen that, in general, high
and low density environment galaxies (the former in groups and the
latter field or isolated) have similar distribution in the Age-[Z/H],
H$\beta$-[MgFe], H$\gamma$-[MgFe] planes, thus suggesting that
dynamical interactions are not a necessary driver for the observed 
variations. If this is true, it confirms 
 Longhetti \etal
(1998a,b, 1999, 2000) who reached the same conclusion by examining the 
line-strength indices of 51
early-type galaxies (21 shell galaxies and 30 members of pairs
displaying clear signatures of interaction) and comparing them with
the reference sample of ``normal'' galaxies by Gonz\'alez
(1993). 

One should be prudent not to overinterpret these results, as the truly 
isolated galaxy statistics in our sample is limited by small number
effects. We could in fact claim to be seeing a different distribution on the
H$\beta$,[MgFe] plane for Fornax as compared with our isolated/field/group 
galaxies in Figure 15.

Our new data seem to be showing a trend for massive (isolated/field/group)
galaxies to be older (Figure 13 and 16-a) and display greater
$\alpha$/Fe values (Figure 11-c).

%%%%%%%%%%%%%%%%%%%%%%%%%%%%%%%%%%%%%%%%%%%%%%%%%%%%%%%%%%%%%%%%%%%%%%%
%%%%%%%%%%%%%%%%%%%% SECTION 8: BROAD PICTURE %%%%%%%%%%%%%%%%%%%%%%%%%
%%%%%%%%%%%%%%%%%%%%%%%%%%%%%%%%%%%%%%%%%%%%%%%%%%%%%%%%%%%%%%%%%%%%%%%

\section{A broad picture of galactic chemical evolution}

Understanding the broad picture of galactic chemical evolution will require 
us to firm up the links between the chemical abundances of predominantly 
active star-forming systems
(i.e.~spirals) and predominantly quiescent ones (i.e.~ellipticals, 
spheroids, lenticulars and bulges). 

While a common elemental yardstick may not exist because of the different elements which we observe directly in each galaxy type, it may be possible to tie the two types together abundances-wise by observing elements in each which share the same nucleosynthesis production site. An example might be oxygen and magnesium. In external spirals oxygen is taken as the metallicity gauge primarily because of its observability. Magnesium, which, like oxygen, is primarily produced in massive stars (Nomoto \etal 1997a,b) may be measurable directly through a calibrated Mg$_2$ index. Then oxygen and magnesium might be linked by assuming a ``cosmic'' Mg/O ratio calibrated locally. Some features in the optical spectra of elliptical galaxies that contain oxygen signatures are G4300 and C$_2$4668, in the near-infrared there is also the 2.3 $\mu$m CO feature. As synthetic spectra and stellar abundances grow more precise, the contribution of specific elements to the indices might be measurable, leading to a much more clear understanding of chemical enrichment and galaxy formation. 

In what follows we have used the H$\alpha$ line information in our spectra of early-type galaxies to estimate the oxygen abundance using the N2 ratio as in Denicol\'o, Terlevich \& Terlevich (2002). We have derived the metallicity [Z/H] for our ellipticals and S0s in Section 4 (this metallicity comes mainly from Mg and Fe abundances), so it is an interesting exercise to compare both [Z/H] metallicity and O abundance estimate for early-type galaxies.

%%%%%%%%%%%%%%%%%%%%%%%%%%%%%%%%%%%%%%%%%%%%%%%%%%%%%%%%%%%%%%%
%%%%%%%%%%%%%%%%% FIGURE 17: BROAD PICTURE %%%%%%%%%%%%%%%%%%%%
%%%%%%%%%%%%%%%%%%%%%%%%%%%%%%%%%%%%%%%%%%%%%%%%%%%%%%%%%%%%%%%
\begin{figure*}
\vspace{12.5 cm}
\includegraphics{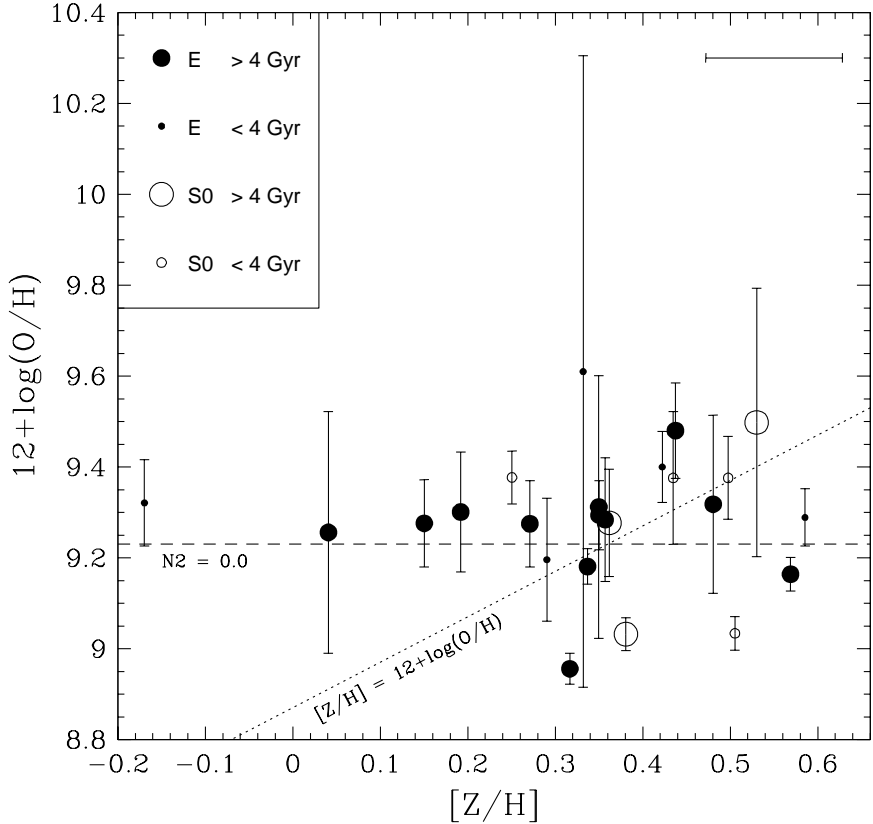}
\caption[Oxygen abundance {\it vs} metallicity Z/H]
{Oxygen abundance {\it vs} metallicity [Z/H]. The oxygen abundance was derived from the \Nii and H$\alpha$ index measurements in the spectra of our early-type galaxies using the N2 calibrator. The [Z/H] metallicity is the result of the interpolation of Thomas \etal (2003) SSP models for measurements in galaxy spectra within the r$_e$/8 aperture. Solid circles indicate elliptical galaxies, open circles are S0s. The size of the points is associated with the galaxy age (ages are from Table 2). The dotted line indicates when 12+log(O/H) is equal to [Z/H]. The dashed line represents 12+log(O/H) when N2 is null.} 
\end{figure*}
%%%%%%%%%%%%%%%%%%%%%%%%%%%%%%%%%%%%%%%%%%%%%%%%%%%%%%%%%%%%%%%

The N2 calibrator is defined as the logarithmic ratio of the emission lines \Nii$\lambda$6584 to H$\alpha$. In the elliptical galaxies, we have measured \Nii and H$\alpha$ as indices. The N2 is the logarithmic ratio of fluxes and not equivalent widths, so we have computed fluxes from the index definitions as stated in eq. 11 of Paper I. We have only used  \Nii and H$\alpha$ indices with negative values  (i.e. lines in emission). The errors in the \Nii and H$\alpha$ indices are derived from repeated measurements. The 12+log(O/H) abundances were  derived from 12+log(O/H) = 9.23($\pm$0.02) + 0.79($\pm$0.03)$\times$N2 
(Denicol\'o, Terlevich \& Terlevich 2002). In Table 4 we show the oxygen abundance and [Z/H] metallicity, for the galaxies with \Nii and H$\alpha$ indices in emission.

%%%%%%%%%%%%%%%%%%%%%%%%%%%%%%%%
%%%% TABLE 4: HA IN ES %%%%%%%%%
%%%%%%%%%%%%%%%%%%%%%%%%%%%%%%%%
%\setcounter{table}{0}
\begin{table*}
\scriptsize
\centering
\small
\hspace*{-1.0cm}
\caption{Oxygen abundance and [Z/H] for early-type galaxies. }
\vspace{0.2 cm}
\begin{minipage}{200mm}
%\begin{tabular}{@{}lrrrrrr@{}} 
\begin{tabular}{ l@{\hspace{.3em}} c@{\hspace{.3em}} c@{\hspace{.3em}} c@{\hspace{.3em}} c@{\hspace{.3em}} c@{\hspace{.3em}} c@{\hspace{.3em}} c@{\hspace{.3em}}}
\hline
\multicolumn{1}{l}{Galaxy} &  
\multicolumn{1}{c}{12+log(O/H)}  &
\multicolumn{1}{c}{error} & 
\multicolumn{1}{l}{[Z/H]} &  
\multicolumn{1}{c}{EW(H$\alpha$)}  &
\multicolumn{1}{c}{er(H$\alpha$)} &
\multicolumn{1}{c}{EW(\Nii)}&
\multicolumn{1}{c}{er(\Nii)}\\ 
\hline
NGC~584  &   9.376   &  0.146   & 0.435   & -0.244    & 0.081   & -0.287    & 0.018 \\
NGC~821  &   9.318   &  0.196   & 0.480   & -0.316    & 0.093   & -0.226    & 0.077\\
NGC~1407 &    9.242  &   0.075  &  0.690  &  -0.531   &  0.085  &  -0.125   &  0.008\\
NGC~1600 &    9.263  &   0.198  &  0.690  &  -0.545   &  0.091  &  -0.233   &  0.099\\
NGC~1700 &    9.376  &   0.091  &  0.498  &  -0.389   &  0.081  &  -0.518   &  0.013\\
NGC~1726 &    9.277  &   0.118  &  0.362  &  -0.800   &  0.166  &  -0.868   &  0.153\\
NGC~2300 &    9.498  &   0.295  &  0.530  &  -0.155   &  0.093  &  -0.380   &  0.122\\
NGC~2418 &    9.164  &   0.037  &  0.569  &  -1.118   &  0.091  &  -1.356   &  0.037\\
NGC~2768 &    9.032  &   0.036  &  0.381  &  -1.317   &  0.097  &  -1.594   &  0.065\\
NGC~2872 &    9.312  &   0.289  &  0.350  &  -0.256   &  0.115  &  -0.162   &  0.080\\
NGC~3156 &    9.321  &   0.095  & -0.170  &  -0.702   &  0.153  &  -0.964   &  0.022\\
NGC~3245 &    8.168  &   0.038  &  0.318  &  -1.806   &  0.083  &  -1.643   &  0.121\\
NGC~3379 &    9.276  &    0.096 &   0.150 &   -0.435  &   0.085 &   -0.198  &   0.020\\
NGC~3599 &    9.400  &   0.078  &  0.422  &  -0.665   &  0.107  &  -1.602   &  0.129\\
NGC~3607 &    9.217  &   0.034  &  0.690  &  -1.022   &  0.080  &  -2.206   &  0.020\\
NGC~3608 &    9.284  &   0.136  &  0.357  &  -0.389   &  0.087  &  -0.190   &  0.042\\
NGC~3665 &    8.401  &   0.024  &  0.690  &  -2.122   &  0.083  &  -1.150   &  0.046\\
NGC~3941 &    9.289  &   0.063  &  0.585  &  -0.663   &  0.084  &  -0.560   &  0.040\\
NGC~4125 &    8.956  &   0.034  &  0.317  &  -1.264   &  0.087  &  -2.644   &  0.098\\
NGC~4261 &    9.480  &   0.105  &  0.437  &  -0.493   &  0.115  &  -1.438   &  0.102\\
NGC~4374 &    9.034  &   0.037  &  0.505  &  -1.308   &  0.107  &  -1.633   &  0.040\\
NGC~4550 &    9.196  &   0.135  &  0.290  &  -1.048   &  0.136  &  -0.268   &  0.076\\
NGC~5322 &    9.372  &   0.092  &  0.690  &  -0.391   &  0.082  &  -0.503   &  0.015\\
NGC~5353 &    9.207  &   0.049  &  0.690  &  -1.029   &  0.090  &  -1.417   &  0.102\\
NGC~5354 &    9.256  &   0.266  &  0.041  &  -0.464   &  0.086  &  -0.144   &  0.084\\
NGC~5444 &    9.301  &   0.132  &  0.192  &  -0.539   &  0.094  &  -0.418   &  0.104\\
NGC~5813 &    9.275  &   0.095  &  0.271  &  -0.784   &  0.092  &  -0.773   &  0.143\\
NGC~5845 &    9.310  &   0.118  &  0.690  &  -0.418   &  0.095  &  -0.305   &  0.045\\
NGC~5846 &    9.181  &   0.039  &  0.337  &  -1.075   &  0.086  &  -1.824   &  0.078\\
NGC~6172 &    9.377  &   0.058  &  0.250  &  -0.729   &  0.092  &  -1.742   &  0.082\\
NGC~7302 &    9.610  &   0.695  &  0.332  &  -0.070   &  0.111  &  -0.286   &  0.060\\
NGC~7585 &    9.523  &   0.102  &  0.690  &  -0.371   &  0.082  &  -1.186   &  0.091\\
NGC~7619 &    9.264  &   0.090  &  0.690  &  -0.635   &  0.084  &  -0.334   &  0.053\\
\hline
\end{tabular}
\vspace{0.2cm}
\end{minipage}
\end{table*}
%%%%%%%%%%%%%%%%%%%%%%%%%%%%%%%%

In Figure 17 we present the 12+log(O/H) {\it vs} [Z/H] diagram. The plot supports substantial oversolar abundances for the interstellar medium (ISM) as measured from 12+log(O/H). If we take the Solar value of 12+log(O/H)$_{\odot}$ = 8.87 (Grevesse \etal 1996), and by definition [Z/H]$_{\odot}$ = 0.0, we draw the dotted line in Figure 17 where 12+log(O/H) = [Z/H]. The distribution of points indicates that most galaxies have Z(ISM) $>$ Z(stars), where we use here ``Z'' to loosely describe metallicity; Z(stars) is taken from [Z/H] and Z(ISM) is inferred from 12+log(O/H). Having Z(ISM) $>$ Z(stars) is consistent with 
our understanding of star formation and evolution: stars cannot have abundances
that are larger than those of the gas from which they formed. The plot in Figure 17 suggests that O/H abundance in the nuclear ISM is similar for all of 
our sample galaxies, irrespective of the luminosity averaged ages and stellar
abundances  inside r$_e$/8. We regard this result as significant.
Denicol\'o, Terlevich \& Terlevich (2002) showed that the relation
N2 to O/H exists for either black-body 
or power-law photoionization. There are several sources of scatter in that 
relation: variations of the photoionization parameter, shape of the
ionizing continuum, differences in secondary production of N, to site
a few. What is important from the present result is the apparent
{\it lack} of scatter suggesting a uniformity of the central ISM composition
irrespective of the mass of the galaxy. If this result is confirmed with a
larger sample of galaxies, one of the consequences of it would be
to rule out galaxy mass related IMF variations.

In other words, we can interpret it as if chemical evolution has
finished in these galaxies, i.e., most stars have already ejected
metals to the ISM and, in general, any currently ongoing star
formation in these galaxies should occur on a small
scale. This is an interesting idea, and could be what
distinguishes most of the chemical characteristics from
early-type to late-type galaxies.

%%%%%%%%%%%%%%%%%%%%%%%%%%%%%%%%%%%%%%%%%%%%%%%%%%%%%%%%%%%%%%%%%%%%%%%
%%%%%%%%%%%%%%%%%%%% SECTION 9: CONCLUSIONS %%%%%%%%%%%%%%%%%%%%%%%%%%%
%%%%%%%%%%%%%%%%%%%%%%%%%%%%%%%%%%%%%%%%%%%%%%%%%%%%%%%%%%%%%%%%%%%%%%%

\section{Conclusions}

We have applied the stellar population model of Thomas, Maraston \& Bender (2003) to estimate ages, metallicities and abundance ratios in a sample of 83 early-type galaxies mostly located in low density regions. 

The stellar population properties derived for each galaxy
correspond to the nuclear r$_e$/8 aperture extraction.  Both
ellipticals and S0s in the sample show a large spread in ages
(from around 1 Gyr to 15 Gyr, but many galaxies also fall out of
the model grids) and metallicities (mostly from 0.0 to 0.67 dex
in [Z/H]). On average, S0
galaxies are slightly younger than the elliptical galaxies by
almost 3 Gyr, and present higher [Z/H] than ellipticals by
approximately 0.10 dex. The average age found for the ellipticals
is 5.8 $\pm$ 0.6 Gyr and the average metallicity is +0.37 $\pm$
0.03 dex. For S0s, the average age is 3.0 $\pm$ 0.6 Gyr and
$<$[Z/H]$>$ = 0.53 $\pm$ 0.04 dex. These averages were performed
excluding the three Virgo cluster galaxies in the sample, and the
uncertainties shown are errors of the mean.

It seems that our elliptical galaxies are in general 3-4 Gyr
younger than E galaxies in the Fornax cluster (Kuntschner 1998,
2000; see Figure 15). In the Fornax cluster the E and S0
galaxies are easily distinguished in terms of age
and metallicity, whereas the Es and S0s from low-density
regions appear to have more homogeneous stellar population
properties; i.e. the differences in age and metallicity are not
so noticeable for Es and S0s of low-density regions than for
Fornax.

 We find that galaxies lie in a plane defined by [Z/H]=0.99log$\sigma_0$$-$0.46log{\it Age}$-$1.60, or in linear terms Z$\propto$$\sigma_0$$\cdot${\it Age}$^{-0.5}$. In this work we are adding more young galaxies to the coverage of the hyperplane defined in Trager \etal (2000b).

Analysing [$\alpha$/Fe] ratios, we can cast some light on the star formation timescale of the galaxies.
More massive (larger $\sigma_0$) and older galaxies present, on average, large [$\alpha$/Fe] values, and therefore, must have undergone shorter star-formation timescales.

Comparing group against field/isolated galaxies, it is still not
clear that environment plays an important role in determining
their stellar population history. In particular, our isolated galaxies show ages
differing by more than 8 Gyr. However, we note that the cluster
elliptical galaxies of Kuntschner (2000) are more uniform in
H$\beta$ (and age) than the group, field and isolated galaxies of
our sample.

Finally, we attempted to draw a broad picture of galactic
chemical evolution by comparing [Z/H] metallicities derived from
SSP models (i.e. stellar metallicity) against log(O/H) abundances
obtained from the N2 ratio of \Nii$\lambda$6584 to H$\alpha$
emission lines (i.e. ISM metallicity). The distribution of points
indicates that most galaxies have Z(ISM) $>$ Z(stars), which is
consistent with chemical evolution expectations.  We interpret it
as if chemical evolution has finished in early-type galaxies,
i.e., most stars have already ejected metals to the ISM and, in
general, any currently ongoing star formation episode in these
galaxies should be on a small scale.

%%%%%%%%%%%%%%%%%%%%%%%%%%%%%%%%%%%%%%%%%%%%
%%%%%%%%     ACKNOWLEDGEMENTS      %%%%%%%%%
%%%%%%%%%%%%%%%%%%%%%%%%%%%%%%%%%%%%%%%%%%%%
\section*{Acknowledgements}

GD would like to thank CNPq-Brazil for the PhD fellowship, INAOE 
for hospitality during the visits to Mexico, Selwyn College and Cambridge
Philosophical Society for financial support.
The authors are grateful to the INAOE Committee for telescope time for 
supporting this project for five consecutive semesters and to 
the staff at Cananea for cheerful logistic help during the 
observations. We thank Am\^ancio Fria\c ca, Alessandro Bressan, Itziar Aretxaga and Olac Fuentes for several interesting discussions and for clarification on the Monte Carlo technique and PCA analysis; Max Pettini and Alfonso Arag\'on-Salamanca for substantial comments to improve this work; and an anonymous 
referee for a careful reading of our manuscript and for suggestions leading 
to a clearer paper.
RJT and ET acknowledge Mexican Research Council Project grants 40018-A-1 and E32186, respectively.

%%%%%%%%%%%%%%%%%%%%%%%%%%%%%%%%%%%%%%%
%%%%%%       BIBLIOGRAPHY       %%%%%%%
%%%%%%%%%%%%%%%%%%%%%%%%%%%%%%%%%%%%%%%

%%%%%%%%%%%%%%%%%%%%%%%%%%%%%%%%%%%%%%%%%%%%%%%%%%%%%%%%%%%%%%%%%%%%%%%%%%%%%%%%%%%%%%%%%%%%%%%%
%%%%%%%%%%%%%%%%%%%%%%%%%%%%%%%%%%%%%%%%%%%%%%%%%%%%%%%%%%%%%%%%%%%%%%%%%%%%%%%%%%%%%%%%%%%%%%%%
%%%%%%%%%%%%%%%%%%%%%%%%%%%%%%%%%%%%%%%%%%%%%%%%%%%%%%%%%%%%%%%%%%%%%%%%%%%%%%%%%%%%%%%%%%%%%%%%
%%%%%%%%%%%%%%%%%%%%%%%%%%%%%%%%%%%%%%%%%%%%%%%%%%%%%%%%%%%%%%%%%%%%%%%%%%%%%%%%%%%%%%%%%%%%%%%%
%%%%%%%%%%%%%%%%%%%%%%%%%%%%%%%%%%%%%%%%%%%%%%%%%%%%%%%%%%%%%%%%%%%%%%%%%%%%%%%%%%%%%%%%%%%%%%%%
%%%%%%%%%%%%%%%%%%%%%%%%%%%%%%%%%%%%%%%%%%%%%%%%%%%%%%%%%%%%%%%%%%%%%%%%%%%%%%%%%%%%%%%%%%%%%%%%
%%%%%%%%%%%%%%%%%%%%%%%%%%%%%%%%%%%%%%%%%%%%%%%%%%%%%%%%%%%%%%%%%%%%%%%%%%%%%%%%%%%%%%%%%%%%%%%%
%%%%%%%%%%%%%%%%%%%%%%%%%%%%%%%%%%%%%%%%%%%%%%%%%%%%%%%%%%%%%%%%%%%%%%%%%%%%%%%%%%%%%%%%%%%%%%%%

\label{lastpage}
\end{document}